\documentclass[a4paper, 11pt]{article}

\usepackage[english]{babel}
\usepackage{bigstrut}
\usepackage{color}
\usepackage{amsmath}
\usepackage{amsfonts}
\usepackage{amssymb}
\usepackage{algorithm}
\usepackage{verbatim}
\usepackage{tabularx}
\usepackage{mathtools}
\usepackage{blkarray}
\usepackage[utf8]{inputenc}
\usepackage[T1]{fontenc}
\usepackage{lmodern}
\usepackage{threeparttable}
\usepackage[bottom,marginal]{footmisc}
\usepackage{soul}
\usepackage{bbm}
\usepackage{hyperref}
\usepackage{array}
\usepackage{setspace}
\usepackage{url}
\usepackage{csquotes}
\usepackage{booktabs}
\usepackage{ltablex}
\usepackage{lscape}
\usepackage{rotating}
\usepackage{multirow}
\usepackage{caption}
\usepackage{graphicx}
\usepackage{xcolor}
\usepackage{acronym}
\usepackage{float}
\usepackage{longtable}
\usepackage{chngpage}
\usepackage{adjustbox}
\usepackage{multicol}
\usepackage{pbox}
\usepackage{authblk}
\usepackage{geometry}
\usepackage{placeins}
\usepackage{threeparttable}
\usepackage{booktabs}  % für \toprule, \midrule, \bottomrule
\usepackage{siunitx}   % für \num{...}
\usepackage{enumitem}
\usepackage[backend=biber,style=authoryear-comp,maxcitenames=2,maxbibnames=99, uniquelist=false,uniquename=init,bibencoding=utf8,doi=false,isbn=false,giveninits=true,
,dashed=false,natbib=true]{biblatex}
\numberwithin{table}{section}
\numberwithin{figure}{section}
\usepackage{subfig}

\makeatletter
\renewcommand\subsubsection{%
    \@startsection{subsubsection}{3}{\z@}%
    {-1.8ex \@plus -0.5ex \@minus -0.2ex}%
    {0.8ex \@plus 0.2ex}%
    {\normalfont\normalsize\itshape}%
}
\makeatother

\begin{document}

\newpage
\selectlanguage{english}
\begin{titlepage}
\setcounter{page}{0}

%\title{AI Substitution of Entry-Level Work \\ Evidence From a Survey Experiment with Managers}
%\title{Task Complexity and Firms' Willingness to Replace Entry-Level Labour with AI: A Survey Experiment with Managers in Germany}
\title{Will AI Replace Juniors? \\ How Task Complexity Reduces Automation Risk}
\author[1,2,3]{Vincent Rost \thanks{\texttt{Corresponding Author; Vincent.Rost@zew.de}.}}
\author[3,4,5]{Fabian Stephany}

\affil[1]{ZEW Mannheim, Germany} 
\affil[2]{Justus Liebig University Giessen, Germany} 
\affil[3]{Oxford Internet Institute, United Kingdom}
\affil[4]{Institute for New Economic Thinking, United Kingdom}
\affil[5]{Bruegel, Belgium}

\date{\today}
%\date{January 2026}
\maketitle

%\centering \textit{\textbf{(Preliminary results - please do not circulate!)}}

\begin{abstract}

How does task complexity shape firms’ willingness to substitute entry-level labour with artificial intelligence (AI) tools? We study this question using a pre-registered survey experiment with managers from 1,435 German firms in the information economy and manufacturing industry. Managers are randomly assigned to consider a long lasting need for additional capacity involving either well-structured and standardised tasks, or more complex tasks that require professional judgement, contextual knowledge and independent problem-solving. Relative to standardised tasks, complex tasks reduce willingness to use AI rather than hire an entry-level employee by 0.54 points on a five-point scale and increase the importance attached to human oversight by 0.15 points. Further, pre-treatment responses indicate that managers expect the availability of AI applications to reduce hiring of both entry-level employees and experienced workers, but significantly more strongly for entry-level workers.
%The results are also relevant for fiscal policy because changes in entry-level hiring and labour market composition may affect tax and social-security revenues as well as public spending on training and labour market support.

\vspace{1cm}

\textit{The work was made possible with the support of a scholarship from the German Academic Exchange Service (DAAD).}
\end{abstract}

\vspace{2cm}
\begin{itemize}
\item[] \textbf{Keywords: AI, labour market, survey experiment, managerial decisions, \mbox{hiring decisions}}
\vspace{-0.25cm}
\end{itemize}

\thispagestyle{empty}
\end{titlepage}
\pagestyle{plain}
\setcounter{page}{1}
\pagenumbering{arabic}
\newpage

\doublespacing

\section{Introduction}

The diffusion of artificial intelligence (AI) and generative AI in particular has renewed concerns about technological displacement and employment. Yet, the emerging debate increasingly suggests that the consequences of the availability of AI applications are unlikely to be uniform across the workforce. Instead of only asking whether AI will destroy or create jobs, a more relevant question is which employees are the most vulnerable, i.e. most exposed to potential substitution, and why. One emerging concern is that AI might disproportionately affect workers at the beginning of their careers, i.e. entry-level employees. 

Recent evidence provides some support for this idea, but remains contested. Hiring and employment have decreased among junior workers in several settings \citep{brynjolfsson-et-al-2026, lodefalk-et-al-2026, tucker-2026}. For instance,  \citet{brynjolfsson-et-al-2026} report that employment among young workers (22--25 years old) in highly AI-exposed occupations has declined substantially compared to those in less AI-exposed occupations, while there is no comparable gap for experienced workers. However, other studies find less strong evidence for AI's impact on junior workers. \citet{lambert-schindler-2026} show that AI exposure is strongly correlated with working from home (WFH) exposure and that the association between AI and the decline in junior hiring attenuates when WFH and AI are both considered. \citet{fairlie-wu-2026} find no significant relative increase in unemployment among recent US college graduates. Similarly, evidence from Germany, is less clear: \citet{arntz2026_iab} find a negative but statistically insignificant association between AI exposure and employment among young workers (aged 20--29) when accounting for WFH exposure and the current economic recession which they see as the main driver of this development.

Thus, whether AI availability systematically reduces demand for junior workers and why it might do so, remains unresolved.

There are several mechanisms that could create such an effect. AI might disproportionately increase the productivity of junior workers, which would allow firms to produce the same output with fewer employees \citep{brynjolfsson-et-al-2025}. Alternatively, AI might be especially capable of automating the type of tasks that is typically done by entry-level workers, consistent with task-based theories of technological change \citep{acemoglu-restrepo-2019, autor-2013, hampole-et-al-2025}. Moreover, AI might lead to a "seniorisation" of entry-level work: When AI is used to automate simple activities, juniors might become responsible for tasks that require judgement, problem-solving and communication \citep{westby-et-al-2026}. Finally, AI applications might change organisational complementarities, allowing experienced employees to supervise AI tools instead of teams of entry-level workers. These mechanisms imply different transformations of entry-level work and career progression.

In this paper, we focus on one central dimension that can help distinguish between these explanations: task complexity. We conduct a preregistered survey experiment with 1,435 managers from German firms in the information economy and the manufacturing industry. Managers are randomly assigned to one of two conditions: Both conditions describe a scenario in which a long lasting need for additional capacity is assumed, but for different types of tasks. One condition describes the tasks in question as standardised and clearly structured entry-level tasks, whereas the other condition refers to entry-level tasks that require professional judgment, contextual knowledge, and independent problem-solving. We then ask whether they would use AI instead of hiring an additional entry-level employee and how important they consider human oversight of AI-generated output for these tasks. 

We find that task complexity significantly affects managers' willingness to use AI instead of human entry-level labour as well as their demand for human oversight. For more complex tasks, managers' willingness to use AI applications instead of hiring an entry-level worker falls by 0.54 points on a five-point Likert scale. At the same time, task complexity significantly increases the importance managers attach to human oversight of AI-generated outputs.

Complementary, non-experimental evidence on managers' hiring expectations over the next two years reveals that they anticipate AI to reduce their hiring activities for both junior and experienced employees. This expected decrease in hirings is significantly stronger for entry-level employees. However, as hires of experienced workers are expected to decrease as well, this adjustment is not an absolute expansion of experienced employment but a relative shift in labour demand towards experience.

Our findings contribute to the literature of AI's impact on labour markets and junior employment by focussing on the potential mechanisms of managerial decisions. The junior status alone does not determine their substitutability; what junior workers do matters.

This complements evidence that AI simultaneously substitutes for some human capabilities while increasing demand for complementary ones \citep{makela-et-al-2026}, and that the composition and complexity of workers' skills shape their resilience to technological change \citep{carpanelli-et-al-2026}. Further, our findings suggest that AI may particularly automate less complex tasks and standardised work through which workers have traditionally entered the workforce, while human oversight and involvement will be preserved where judgement and contextual knowledge is required. This raises a longer-term question beyond immediate employment effects: If AI is used to perform the tasks through which junior workers traditionally acquired expertise, how will firms develop the experienced workers they continue to value?

The remainder of this paper is structured as follows: Section 2 elaborates on the conceptual background and derives our hypotheses; Section 3 explains the data source and experimental design; the empirical strategy is explained in Section 4. We present our results in Section 5 and provide additional analyses and robustness checks in Section 6. Section 7 provides a discussion, before Section 8 concludes.

\section{Conceptual Background and Hypotheses}

AI does not uniformly substitute human capabilities. Its diffusion can simultaneously reduce demand for substitutable skills while increasing demand for complementary ones \citep{makela-et-al-2026}. One prominent concern is that these dynamics may disproportionately affect workers who are at the beginning of their careers. Recent studies document declining employment or hiring among young workers in occupations with high AI exposure in the United States and Sweden \citep{brynjolfsson-et-al-2026, lodefalk-et-al-2026, tucker-2026}. However, the evidence remains contested. The apparent effect of AI becomes considerably weaker when accounting for WFH and remote work in some settings \citep{lambert-schindler-2026}, while other studies find no distinct post-LLM decline in junior hiring or higher rates of unemployment \citep{fairlie-wu-2026, oecd-2026}. Using data from 110 economies, \citet{ando2026} detect a relative decline in online job postings for occupations with higher AI exposure after the release of ChatGPT. However, they do not find a statistically significantly stronger decline for junior vacancies compared to senior vacancies. Hence, both the extent to which AI reduces demand for junior workers and, equally importantly, why it might do so, remain unresolved. We distinguish four mechanisms that could lead to a decline in demand for junior workers while implying fundamentally different transformations of work.

\subsection{Productivity Paradox}

First, a new kind of productivity paradox might arise if AI disproportionately increases productivity of junior workers. Generative AI produces substantial productivity gains for knowledge work \citep{dellacqua-et-al-2023, noy-zhang-2023}, with some of the largest gains for less experienced workers \citep{brynjolfsson-et-al-2025}. However, an increase in worker productivity does not necessarily lead to increased labour demand. Instead, firms might require fewer (junior) employees to produce the same output. Thus, AI can complement junior workers on an individual level and simultaneously reduce firms' demand for them \citep{acemoglu-restrepo-2019, hampole-et-al-2025}.

\subsection{Task Substitution}

Second, AI tools might lead to task substitution. Technologies substitute tasks rather than entire occupations \citep{acemoglu-restrepo-2019, autor-2013}, and as junior employees are typically responsible for simpler tasks such as retrieving information, drafting and summarising, basic analyses, translations or simple coding, their tasks have the largest overlap with current AI capabilities. Occupations with AI-exposed task bundles are associated with a decline in labour demand \citep{hampole-et-al-2025}. At the skill level, AI reduces demand for skills it can replace, but increases demand for skills that complement it \citep{makela-et-al-2026}. Junior workers may therefore be especially vulnerable not because they are junior per se, but because of the tasks they perform. Given AI's “jagged technological frontier” \citep{dellacqua-et-al-2023}, this mechanism predicts substantially weaker substitution risks where junior workers perform more complex tasks.

\subsection{Seniorisation}

Third, AI might lead to a seniorisation of junior work. Instead of eliminating entry-level positions, firms might shift their tasks and content towards tasks that were formerly associated with more experienced workers and require skills like independent problem solving, critical judgement and evaluation. For instance, \citet{westby-et-al-2026} find a decline in junior software vacancies compared to senior software vacancies after ChatGPT was introduced. Simultaneously, the remaining junior roles asked for higher levels of experience, problem solving and communication skills. Similarly, AI-intensive jobs increasingly reward complementary human capabilities such as analytical thinking and resilience \citep{makela-et-al-2026}. Consequently, AI might shape which tasks, skills and capabilities are associated with entry-level positions, increasing demand for junior employees who already possess capabilities conventionally associated with more senior roles.

\subsection{AI Complementarity}

Fourth, AI may create a new complementarity between senior employees and AI. Knowledge-intensive work has traditionally combined junior employees who execute tasks with senior employees who provide expertise, judgment, and supervision. AI applications have the potential to change this relationship by allowing experienced workers to delegate task execution to AI and supervise these outputs instead. Consequently, the production complementarity may shift from senior-junior to senior-AI. This is in line with recent evidence that automating some tasks can increase the value of human input into the remaining tasks \citep{hampole-et-al-2025}, particularly capabilities complementary to AI \citep{makela-et-al-2026}, and that human expertise becomes especially important when tasks extend beyond the current capabilities of AI tools \citep{dellacqua-et-al-2023}.

\subsection{Investigating Task Complexity}

In order to distinguish those mechanisms, this paper moves beyond occupation titles and seniority labels towards the nature and content of the tasks of the underlying work. Recent evidence shows that workers' skill portfolios can predict their career outcomes and resilience: Workers who have higher degrees of specialisation tend to work in higher-paying occupations, while employees who have skills in several fields are more likely to acquire new skills and move between occupations. Workers with broader skills are also more likely to move into occupations with less automation exposure \citep{carpanelli-et-al-2026}. Therefore, task complexity may provide an important factor of distinguishing mechanisms that otherwise produce observationally similar declines in junior employment.

The aforementioned mechanisms give rise to different predictions: The productivity mechanism predicts a decline in demand for junior workers irrespective of task complexity, because the relevant distinction is the workers' experience. Task substitution instead predicts that reduced demand for junior labour will be concentrated among simple rather than complex tasks. The seniorisation mechanism predicts increased demand for junior employees who possess skills and capabilities that are traditionally associated with more senior roles. Finally, senior–AI complementarity predicts greater importance of human oversight for more complex tasks, and increased demand for senior workers to provide such supervision.

Together, these mechanisms shift the question from whether AI will replace junior labour to how AI might reorganise tasks, expertise, collaboration and supervision across different levels of seniority. We hypothesise that AI's effect on the demand for junior workers is conditional on the nature and complexity of their tasks.

\subsection{Hypotheses}
\label{subsec:hypotheses}

This implies that the substitutability of entry-level work depends on the content and the nature of the task for which the firm requires additional capacity. Standardised and clearly structured tasks can more readily be specified, delegated to an AI system, and assessed against a pre-defined output. When tasks require contextual knowledge, professional judgement, and independent problem-solving, however, the task is harder to delegate to AI tools. We therefore expect managers to be less willing to replace an additional entry-level worker with AI when the required work is complex:

\medskip
\noindent\textit{H1: Firms are more likely to use AI applications instead of hiring an additional entry-level employee when the required tasks are standardised and clearly structured rather than complex.}
\medskip

Task complexity may also affect how firms intend to use AI when they do deploy it. Outputs from standardised processes are comparatively easy to verify, while errors in context-dependent tasks may be harder to detect and have more severe consequences. We therefore assume that complex work should increase the perceived importance of human oversight of potential AI outputs.

\medskip
\noindent\textit{H2: Firms attach greater importance to human oversight of AI-generated output when the required tasks are complex rather than standardised and clearly structured.}
\medskip

Hypotheses~1 and~2 concern different organisational decisions. The first addresses whether AI is considered an alternative to an additional entry-level hire. The second concerns the degree of human involvement managers consider necessary when AI is used. 

Beyond the experimental scenario -- i.e. before the experimental treatment takes place -- the survey elicits managers' expectations about how the availability of AI tools will affect their firm's hiring decisions over the following two years. Entry-level positions often require standardised or well-structured tasks and activities that can be at least partly, easily automated. If AI allows existing employees to complete these activities more efficiently, firms may require fewer additional junior workers. This leads to the following prediction:

\medskip
\noindent\textit{H3: Managers expect the availability of AI applications to reduce their firm's hiring of entry-level employees over the following two years.}
\medskip

The expected effect on experienced employees is less clear. AI may automate some of their activities as well, but at the same time, expertise and domain knowledge will be needed to oversee and critically judge AI outputs. Our preregistered hypothesis therefore emphasises complementarity between AI applications and experienced\footnote{Our questionnaire defined those as "employees with professional experience, i.e., those with in-depth knowledge in their field"} workers:

\medskip
\noindent\textit{H4: Managers expect the availability of AI applications to increase their firm's hiring of experienced employees over the following two years.}
\medskip

Hypotheses~3 and~4 therefore predict that AI serves as a substitute for junior-level workers, but as a complement for experienced workers. They are evaluated separately against the neutral response category of no expected effect. Unlike Hypotheses~1 and~2, these comparisons are descriptive. The questions are causal in their wording, as managers are directly asked how the availability of AI will affect their firm's future hiring. This causal judgement is, however, supplied by the respondent, rather than identified through experimental variation. The estimates therefore capture managers' perceived causal effects of AI availability, not the causal effect of AI on realised employment. We additionally compare the two expected hiring effects within firms. 

These four hypotheses constitute the confirmatory hypotheses specified in the preregistration.\footnote{The preregistration is available at \url{https://aspredicted.org/ix9za2.pdf}.}

\section{Data and Experimental Design}

\subsection{Survey and Sample}
\label{subsec:survey_sample}

The firm survey 'Konjunkturumfrage Informationswirtschaft' that is conducted every quarter by ZEW (Leibniz Centre for European Economic Research, Mannheim, Germany) is the data source for this research project. In its current format, it has been running since 2011 and produces between 1,000 and 1,500 responses each quarter. Originally, this survey only covered firms that belong to the information economy (e.g. ICT hardware, ICT services and knowledge-intensive services such as media agencies, tax consultants etc.) and have at least 5 employees. Recently, however, the scope of the survey was extended as it now also regularly includes firms in the manufacturing industry (also with at least 5 employees), which also allows comparisons across both industries. 

The questionnaires are sent by post directly to the CEOs of the firms. Even though we can't fully rule out that the paper questionnaire is handled by other employees within the company, we are fairly confident that the vast majority of our respondents has substantial decision power in the firm. In a previous wave of the panel we explicitly asked the respondent for their position within the firm: Around 73\% of the respondents were CEOs or members of the executive board, 17\% reported to be department heads (or holding higher managing positions), while less than 3 percent were team or project leaders. Only 7 percent of respondents did not hold any management function. Even though these numbers stem from another wave of the survey, we are confident to indeed reach CEOs and decision-makers of German firms with our instrument. 

After receiving the questionnaire, which consists of only one page with easy-to-answer items, they are presented with several ways to respond: They can either use the provided paper questionnaire and send the filled-in version back to ZEW via post or scan it and send it via e-mail or fax machine. Alternatively, they can use an online log-in to answer the questions using an online form. In total, 1,059 respondents (73.8 per cent) participated online and 376 (26.2 per cent) used an offline mode.

The data for this paper come from wave 129 of the aforementioned firm survey, which was fielded in mid 2026. Responses using the online form were recorded between 19 June and 5 August 2026. A small number of filled in paper questionnaires were sent to us after the online field period had already ended, but were also retained, with the final accepted questionnaire entered on 26 August 2026. After a set of recurring questions regarding current business conditions, expectations and development of the firm's headcount, each survey wave contains questions on specific issues revolving around the topics of digitalisation, innovation and the organisation of work. This wave focused on firms' use of artificial intelligence, hiring behaviour and their demand for entry-level labour.

After excluding firms that do not belong to the pre-defined industry or have fewer than five employees, the final sample consists of 1,435 observations. Of these, 768 (53.5\%) belong to the information economy, 667 (46.5\%) operate in the manufacturing industry. Firms are distributed fairly evenly across the three size classes used in the analysis:\footnote{This is not representative for the size distribution of German firms; pooled results should therefore be treated with caution, as large firms are overrepresented in our sample} 448 (31.2\%) have 5--19 employees, 550 (38.3\%) have 20--99 employees, and 437 (30.5\%) have at least 100 employees. The median firm has 38 employees, whereas the mean is 172.9.

Regarding personal characteristics, among respondents identifying as either female or male, 24.7\% are female and 75.3\% are male.\footnote{Two respondents selected ``other'' as their gender, 49 selected ``prefer not to answer'', and 46 did not provide a gender response.}  As the survey directly addresses managers and employees with high levels of decision power, the age distribution is shifted towards the upper end of the scale: 71.3\% of respondents are at least 50 years old.\footnote{Forty respondents did not report their age category.} 

Interestingly, the sample was already relatively familiar with AI applications. The firms report using an average of 2.2 of the six AI application categories listed in the questionnaire, and 67.3 per cent of respondents use AI applications either daily or several times per week. The experiment therefore elicits responses and hiring expectations from managers with often direct exposure to AI. Table~\ref{tab:sample_characteristics} reports the composition of the sample along categorical (Panel A) and numerical (Panel B) characteristics.\footnote{The analytical samples for the  outcomes are documented in Appendix Table~\ref{tab:analytical_samples}.}

\begin{table}[htbp]
    \centering
    \begin{threeparttable}
        
        \footnotesize
        \begin{tabular*}{\textwidth}{@{\extracolsep{\fill}}llrrrrr}
\toprule
Characteristic & Category & $N$ & Percent & Mean & SD & Median \\
\midrule
\multicolumn{7}{l}{\textit{Panel A: Sample composition}} \\
\addlinespace[0.3em]
Industry & ICT sector &  216 & 15.1 &  &  &  \\
 & Media services &   80 &  5.6 &  &  &  \\
 & Knowledge-intensive services &  472 & 32.9 &  &  &  \\
 & Manufacturing &  667 & 46.5 &  &  &  \\
Firm size & 5--19 employees &  448 & 31.2 &  &  &  \\
 & 20--99 employees &  550 & 38.3 &  &  &  \\
 & 100+ employees &  437 & 30.5 &  &  &  \\
Survey mode & Offline &  376 & 26.2 &  &  &  \\
 & Online & 1059 & 73.8 &  &  &  \\
Respondent characteristics & Female &  330 & 24.7 &  &  &  \\
 & Aged 50 or older &  995 & 71.3 &  &  &  \\
AI experience & Frequent personal AI use &  966 & 67.3 &  &  &  \\
Previous entry-level hiring & Vocational qualification &  760 & 53.0 &  &  &  \\
 & University degree &  655 & 45.6 &  &  &  \\
\midrule
\multicolumn{7}{l}{\textit{Panel B: Numeric characteristics}} \\
\addlinespace[0.3em]
Employees in 2025 &  & 1403 &  & 172.9 & 879.5 & 38.0 \\
Employees with a university degree (\%) &  & 1316 &  & 31.5 & 31.1 & 20.0 \\
AI applications used &  & 1430 &  & 2.2 & 1.5 & 2.0 \\
Hiring difficulty: vocational &  & 1107 &  & 3.0 & 1.3 & 3.0 \\
Hiring difficulty: university &  & 1039 &  & 2.9 & 1.3 & 3.0 \\
\bottomrule
\end{tabular*}

        \caption{Sample characteristics}
        \label{tab:sample_characteristics}
        
        \begin{tablenotes}[flushleft]
            \footnotesize
            \item \textit{Notes:} The table describes the characteristics of the eligible sample. For the gender question, respondents were given the outside option to not disclose this information. These were excluded for calculating the shares of female respondents. Frequent personal AI use is defined as using AI tools at least several times per week. Previous entry-level hiring indicates whether a firm hired at least one entry-level employee during the previous two years, differentiated for their qualification. The number of AI applications could range between 0 and 6 of the presented options. Hiring difficulties are measured on a 5-point Likert scale and are only reported for firms that indicated a need for the respective group of employees. 
            %\item \textit{Notes:} The table describes the eligible sample of 1,435 firms. Except for respondent gender and age, percentages in Panel A are calculated relative to the full eligible sample. The share of female respondents is calculated among respondents who identified as either female or male (\(N=1{,}338\)); two respondents selected another gender category, 49 preferred not to answer, and 46 did not provide a gender response. The share of respondents aged 50 or older is calculated among respondents reporting a valid age category (\(N=1{,}395\)); 40 respondents did not report their age. For binary characteristics, \(N\) reports the number of observations meeting the stated condition; the remaining observations belong to the respective reference category or, where applicable, have missing information. Frequent personal AI use is defined as using AI tools daily or several times per week. Previous entry-level hiring indicates that the firm hired at least one entry-level employee with the respective qualification during the previous two years. The number of AI applications ranges from zero to six. Hiring difficulties are measured on a five-point scale, with higher values indicating greater difficulty, and are reported only for firms indicating a need for the respective qualification group.
        \end{tablenotes}
        
    \end{threeparttable}
\end{table}

\FloatBarrier
\subsection{Experimental Design and Randomisation}

In order to causally identify the effect of task complexity on our outcome measures, each firm is randomly assigned to one of two groups. Both groups received largely identical questionnaires that only differed in the task complexity for the given scenario. Respondents were asked to imagine a scenario in which they needed additional capacities to fulfil a task and that this additional need would persist long-term. The experimental variation alters the level of complexity for the given task. In the high-complexity group, the relevant task was described as being suitable for entry-level workers, "but require professional judgement, contextual knowledge and independent problem-solving". Additionally, we provided examples to underline the type of tasks we were referring to. 

The other group received a slightly different setting. The scenario was identical (long-term need for additional capacity), but the tasks in question were less complex. They were defined as suitable for entry-level workers and characterised as standardised and clearly structured. Accordingly, we provided different examples to clarify the kind of task in the low-complexity setting. 

The complete wording can be seen in Table~\ref{tab:experimental_design}.

\bigskip

\begin{table}[htbp]
    \centering
    \captionsetup{position=bottom,skip=0.8em}

    \small
    \begin{tabular}{
        @{}
        >{\raggedright\arraybackslash}p{0.485\textwidth}
        @{\hspace{0.03\textwidth}}
        >{\raggedright\arraybackslash}p{0.485\textwidth}
        @{}
    }
        \toprule
        \textbf{High task complexity}
        (\(n=705\))
        &
        \textbf{Low task complexity}
        (\(n=730\))
        \\
        \midrule

        Please imagine the following situation: Your company requires additional capacity in the long term for tasks that are typically performed by entry-level employees but require \textbf{professional judgement, contextual knowledge, and independent problem-solving}. Examples include interpreting information, preparing individualised recommendations, or handling more complex customer enquiries.
        &
        Please imagine the following situation: Your company requires additional capacity in the long term for \textbf{standardised, clearly structured tasks} that are typically performed by entry-level employees. Examples include data maintenance, simple research, standardised analyses, or handling recurring customer enquiries.
        \\

        \bottomrule
    \end{tabular}

    \caption{Experimental task-complexity manipulation}
    \label{tab:experimental_design}

    %\vspace{-0.4em}
    \begin{minipage}{\textwidth}
        \footnotesize
        \textit{Notes:} Respondents were randomly assigned to one of the two scenarios. The questionnaire versions differed only in the task description shown in the table; the subsequent outcome questions and response scales were identical across treatment groups. The table presents English translations of the original German questionnaire.
    \end{minipage}
\end{table}

\bigskip

% zu informell?? As we send out paper questionnaires by post and only a small fraction of all contacted firms respond, randomisation is not trivial. Whereas we could easily achieve an equal 50:50 split in a pure online survey, this is much harder for a printed questionnaire and modest response rates. Hence, our randomisation process includes several steps, i.e. we first randomly assign those firms that took part in one of the previous waves as they have the highest chance to take part again. Moreover, the randomisation is conditioned on sectors and firm sizes. In our case, this randomisation process worked well with 705 responding firms (49.1\%) in the high-complexity setting and 730 firms (50.9\%) in the low-complexity group. 

The random assignment to one of the two groups took place at the firm level before the questionnaires were printed and distributed. As the randomly varied scenarios are part of the printed questionnaire, this timing is necessary and makes the randomisation process somewhat more difficult compared to a pure online survey where participants can be randomly assigned to a treatment condition in the moment of their participation. In total, 11,748 firms were contacted for this wave of the survey. To increase the chance of a well-balanced randomisation, the randomisation process was done in two stages: First, all firms that took part in at least one of the two previous waves of the survey (i.e. firms with a higher likelihood of participating again) were randomly assigned to one of the two complexity conditions. In a second step, the remainder of firms in our initial sample is randomly assigned. Before the questionnaires were sent out, the distributions of firm size and sector were checked across both groups. Hence, the randomisation procedure can be understood as a conditioned random assignment based on recent survey participation. In the final sample of firms that participated in the current survey wave, 705 respondents received the high-complexity questionnaire and 730 received the low-complexity questionnaire.

Table~\ref{tab:balance} presents the balance between both groups on all relevant characteristics.  Overall, the groups are well balanced. Among the individual comparisons, only the share of firms with at least 100 employees differs significantly at the 5\% level: Within the high complexity group, 32.9\% of firms have at least 100 employees, whereas this share amounts to 28.1\% in the low-complexity group, a difference of 4.8 percentage points ($p=0.047$). Given the number of direct comparisons, one marginally significant difference may arise by chance even when the randomisation was successful.

Moreover, a formal joint Wald test does not reject equality of the observed characteristics across the treatment groups ($\chi^2(12)=9.54$, $p=0.656$). The statistically significant difference in firm size therefore provides no indication of systematic imbalance between the groups. Further, industry and firm-size fixed effects are included in the reported model specifications, so any observed differences cannot be driven by differences in firm size composition. %The treatment estimates remain very similar after these controls are included, showing that the main results are not driven by the somewhat larger share of large firms in the high-complexity group.

\begin{table}[htbp]
    \centering
    
\begin{tabular}{lrrrrr}
\toprule
Variable & Mean High & Mean Low & Diff. & t-stat. & p-value\\
\midrule
Information economy & 0.53 & 0.54 & -0.01 & -0.56 & 0.5743\\
Number of employees & 207.64 & 139.24 & 68.40 & 1.44 & 0.1497\\
5–19 employees & 0.30 & 0.33 & -0.03 & -1.15 & 0.2500\\
20–99 employees & 0.37 & 0.39 & -0.02 & -0.78 & 0.4339\\
100+ employees & 0.33 & 0.28 & 0.05 & 1.99 & 0.0473\\
\addlinespace
Online response & 0.73 & 0.74 & -0.01 & -0.32 & 0.7489\\
Female respondents & 0.25 & 0.25 & 0.00 & -0.10 & 0.9185\\
Respondents aged 50+ & 0.71 & 0.72 & -0.01 & -0.29 & 0.7724\\
AI applications used & 2.24 & 2.17 & 0.06 & 0.80 & 0.4233\\
Frequent personal AI use & 0.68 & 0.68 & 0.00 & 0.09 & 0.9285\\
\addlinespace
Any entry-level hires: vocational & 0.56 & 0.53 & 0.03 & 1.04 & 0.2983\\
Any entry-level hires: university & 0.50 & 0.45 & 0.05 & 1.85 & 0.0644\\
Hiring difficulty: vocational & 3.02 & 3.02 & 0.00 & 0.06 & 0.9554\\
Hiring difficulty: university & 2.86 & 2.91 & -0.05 & -0.62 & 0.5369\\
\bottomrule
\end{tabular}
    \caption{Balance between the high- and low-task-complexity treatment groups}
    \label{tab:balance}

    \vspace{0.2cm}
    \begin{minipage}{\textwidth}
        \footnotesize
        \textit{Notes:} The table reports treatment-group means, their difference (high minus low task complexity), and two-sided tests of equality. A joint Wald test of the pre-treatment covariates does not reject joint equality between the treatment groups (\(p=0.656\)).
    \end{minipage}
\end{table}

\subsection{Outcome Variables and Covariates}
\label{subsec:variables}

The two main outcomes of interest were elicited immediately after the experimentally varied scenario and referred explicitly to the type of task described in the respective treatment condition.

\subsubsection{AI substitution}

The first post-treatment outcome measures the firm's stated likelihood of using AI applications instead of hiring an additional entry-level employee to fulfill this capacity demand. The exact wording was:

\begin{quote}
\textit{How likely would your company be to use AI applications for such tasks instead of hiring an additional entry-level employee?}
\end{quote}

Responses were recorded on a five-point Likert scale ranging from 1 ("very unlikely") to 5 ("very likely"). In addition to this scale, respondents were offered an outside option: "We already do this". Hence, higher scale values indicate a greater willingness to substitute additional entry-level hire by AI tools.

The main analysis uses the five-point response whenever a scale category is available. This includes the small number of questionnaires on which both a scale category and the outside option were selected. A valid scale response is available for 1,371 firms. For robustness checks, these ambiguous responses were excluded, as reported in Section~\ref{subsec:robustness}.  

\subsubsection{Human oversight}

The second main outcome measures the importance managers assign to human oversight of potential AI-generated output for the respective task:

\begin{quote}
\textit{How important would it be to you that AI-generated results for such tasks are critically reviewed by a human?}
\end{quote}

The five-point scale ranges from 1 (not at all important)  to 5 (very important). The corresponding outside option was "We generally do not use AI". Thus, higher values indicate greater perceived importance of human oversight for AI outputs. We use all available responses on the five-point scale. Again, the main specification includes the small number of respondents who selected both a scale value and the outside option. Overall, 1,359 firms provided a valid scale response. For robustness checks (see Section~\ref{subsec:robustness}) these were excluded.

%Appendix Table~\ref{tab:response_patterns} documents the response patterns for both main outcomes. As a robustness check, Section~\ref{subsec:robustness} reports estimates that exclude respondents who selected both a scale value and the corresponding outside option.

\subsubsection{Expected hiring effects}

Before the experimentally varied scenario, respondents were asked how they expect the availability of AI applications to affect their firm's hiring behaviour over the following two years. The question asked separately for the expected change in hiring behaviour referring to entry-level employees and to employees with professional experience, defined in the questionnaire as workers with established knowledge in their field.

Both outcomes are measured on a fully verbalised five-point scale: 1, ``substantially decrease''; 2, ``slightly decrease''; 3, ``no effect''; 4, ``slightly increase''; and 5, ``substantially increase''.
Additionally, respondents could select the outside option ``generally no need'' for the respective employee group. Because this outside category has no natural position on the directional scale, the tests of Hypotheses~3 and~4 use only responses between 1 and 5. Valid scale responses are available for 1,247 firms for entry-level hiring and 1,291 firms for experienced hiring; 1,230 firms provided valid responses for both questions.

\subsubsection{Covariates}

The adjusted specifications use variables on firm and respondent characteristics. For firm characteristics, four industry categories and three firm-size classes are used (as listed in Table~\ref{tab:sample_characteristics}). Respondent characteristics comprise indicators for a female respondent and for a respondent aged 50 or older.

Moreover, prior AI experience is measured, both on a firm as well as on a personal level. Firm-level experience is the number of AI application categories\footnote{See Appendix~\ref{app:ai_experience}, Q1, for the complete list of response options.} currently used by the firm, ranging from zero to six. Personal AI use was elicited by a question regarding the frequency the respondent personally uses AI applications. Frequent personal AI use equals one if the respondent reports using AI applications daily or several times per week.

Subsequently, respondents were asked about their hiring behaviour and difficulties, separately for entry-level employees with a vocational qualification, and for entry-level employees with a university degree. After eliciting the number of respective hires over the past two years, they were asked to assess the difficulty in finding and hiring entry-level employees of both kinds, with the outside option of stating that they generally don't need the respective kind of entry-level employee. The scale ranges from 1, ``no difficulty'', to 5, ``great difficulty''. In the regression models, recruitment difficulty is set to zero for firms reporting no need, while separate need indicators distinguish these firms from those reporting a need and low recruitment difficulty.

Two binary indicators capture whether the firm hired at least one entry-level employee with a vocational qualification or a university degree during the previous two years. 

The preregistration specified the perceived importance of entry-level employees for the firm’s talent pipeline as a pre-treatment measure and a control variable. Subsequently, we moved this question to the post-treatment set of questions. This deviation from the preregistered intention was motivated by the concern that asking about the importance of the firms' internal talent pipeline could prime respondents to focus on the long-term consequences of their hiring decisions, which might have altered their response on the task complexity treatment. As the question was asked after the treatment, i.e. could be affected by the treatment itself, we did not include it as a control variable. 

Appendix Table~\ref{tab:analytical_samples} reports the number of observations and treatment allocation in each analytical sample.

\section{Empirical Strategy}
\label{sec:empirical_strategy}

\subsection{Main Specifications}
\label{subsec:main_specifications}

As we use experimental data and randomisation worked well, we estimate the following OLS specification\footnote{All OLS models use heteroskedasticity-robust standard errors (HC3)} for all experimental outcomes:

\begin{equation}
    \label{eq:baseline_OLS}
    Y_i
    =
    \beta_0
    + \beta_1 \textit{HighComplexity}_i
    + \beta_2' X_i
    + \varepsilon_i .
\end{equation}

where \(Y_i\) denotes either the five-point AI-substitution score or the five-point human-oversight score for firm \(i\). The constant \(\beta_0\) represents the expected outcome in the low-complexity condition (with all included covariates equalling zero or their reference category). The indicator \(\textit{HighComplexity}_i\) equals one if firm \(i\) received the high-complexity scenario and zero if it received the low-complexity scenario. \(\beta_1\) is the main coefficient of interest and captures the average difference in the respective outcome between the high- and low-complexity conditions, i.e. the treatment effect of being in the high complexity setting. The vector \(X_i\) contains the control variables included in the respective specification, while \(\beta_2'\) denotes the corresponding vector of coefficients. Finally, \(\varepsilon_i\) is the error term.

Hypothesis~1 predicts \(\beta_1<0\) for AI substitution, whereas Hypothesis~2 predicts \(\beta_1>0\) for human oversight.

We report four nested specifications. Model~(1) is the baseline model and contains only the treatment indicator as an independent variable. Model~(2) adds industry and firm-size fixed effects, whereas Model~(3) also includes respondent characteristics (gender and age). Model~(4) adds firm and personal AI experience, previous entry-level hiring, and the qualification-specific need and recruitment-difficulty measures described in Section~\ref{subsec:variables}. For all plots and models of our main model specifications we consistently use the set of complete cases used in Model~(4). We thereby make sure that changes across columns can be attributed to covariate adjustment and do not stem from a shift in sample composition. The corresponding common samples contain 1,210 observations for AI substitution and 1,205 observations for the model for human oversight.

The hiring-expectation outcomes are not generated by the experimental manipulation. To evaluate Hypotheses~3 and~4, we test the mean response on the 5-point Likert scale against the neutral scale value of 3. This is done for both types of employees separately. We then compare the two expectations within the 1,230 firms that provided valid responses for both items. This paired within-respondent comparison tests whether the expected change ibn future hiring behaviour differs between entry-level and experienced hiring. These estimates describe managers' expectations, which are causal in their wording, but not experimental or otherwise causal measurements on our side.

%\subsection{Inference and Sample Definitions}
%\label{subsec:inference}

\section{Results}
\label{sec:results}

\subsection{AI Substitution of Entry-Level Work}
\label{subsec:kisub_results}

% KI-recommendation: nicht von probability sprechen, sondern: Table~\ref{tab:kisub_main} shows the results of the OLS models for managers' stated willingness to use AI applications for the given tasks instead of hiring an entry-level employee.

\begin{figure}[htbp]
    \centering
    \includegraphics[
        width=\textwidth
    ]{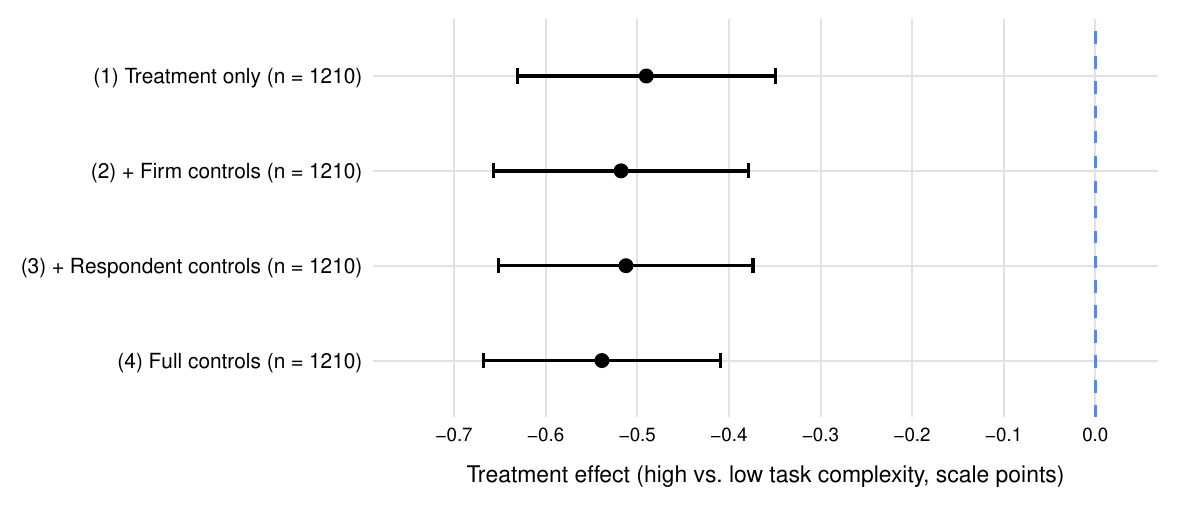}
    \caption{Effects of task complexity on AI substitution}
    \label{fig:kisub_treatment_effects}
    
    \vspace{0.2cm}
    \begin{minipage}{\textwidth}
        \footnotesize
        \textit{Notes:} Points show OLS estimates of the effect of high
        relative to low task complexity on firms' willingness to use AI
        instead of hiring an entry-level employee. Horizontal lines indicate
        95\% confidence intervals based on HC3-robust standard errors.
        All specifications are estimated using the common sample of the
        full model ($N=1{,}210$).
    \end{minipage}
\end{figure}

Table~\ref{tab:kisub_main} shows the results of the aforementioned OLS model with the managers' assessed probability of using AI applications for the given task instead of hiring an entry-level employee. The baseline model in column (1) of Table~\ref{tab:kisub_main} shows a negative estimation for the main coefficient $\beta_1$ that is significant on all conventional levels. Adding firm and respondent characteristics, as well as the pre-specified covariates does not change direction, size or significance of the coefficient estimation. This means that managers are significantly less willing to use AI applications instead of an additional entry-level employee when the task that needs to be done is complex. Thus, the results support Hypothesis~1. If all control variables are accounted for, higher task complexity lowers the AI-substitution score by 0.54 points on the 5-point Likert scale. Figure~\ref{fig:kisub_treatment_effects} plots the four estimates and their 95\% confidence intervals. 
Further, Model~(4) provides additional insights: Higher levels of AI use -- both on the firm and on the personal level -- correspond with higher likelihoods of using AI instead of hiring an additional entry-level employee. Neither coefficient regarding past difficulties in recruitment is statistically significant at the 5\% level. The coefficients on recruitment difficulty and AI use describe conditional associations rather than causal effects.

\bigskip

% AI substitution

\begin{table}[htbp]
    \centering
    
\begin{tabular*}{\textwidth}{@{\extracolsep{\fill}}lcccc}
\toprule
Variable & (1) & (2) & (3) & (4)\\
\midrule
High task complexity & -0.490*** & -0.518*** & -0.512*** & -0.539***\\
 & (0.072) & (0.071) & (0.071) & (0.066)\\
AI applications used &  &  &  & 0.258***\\
 &  &  &  & (0.028)\\
Frequent personal AI use &  &  &  & 0.389***\\
\addlinespace
 &  &  &  & (0.078)\\
Any previous entry-level hires: vocational &  &  &  & -0.054\\
 &  &  &  & (0.084)\\
Any previous entry-level hires: university &  &  &  & 0.104\\
 &  &  &  & (0.082)\\
\addlinespace
Hiring difficulty: vocational &  &  &  & 0.066+\\
 &  &  &  & (0.038)\\
Hiring difficulty: university &  &  &  & -0.052\\
 &  &  &  & (0.040)\\
\midrule
$R^2$ & 0.037 & 0.076 & 0.087 & 0.222\\
\addlinespace
Mean dep. var. (low complexity) & 3.208 & 3.208 & 3.208 & 3.208\\
Industry FE & No & Yes & Yes & Yes\\
Firm size FE & No & Yes & Yes & Yes\\
Respondent controls & No & No & Yes & Yes\\
Need indicators & No & No & No & Yes\\
\addlinespace
Observations & 1210 & 1210 & 1210 & 1210\\
\bottomrule
\end{tabular*}
    \caption{Effects of task complexity on the likelihood of using AI instead of hiring an entry-level employee. HC3-robust standard errors are reported in parentheses. Significance levels: + \(p<0.10\), * \(p<0.05\), ** \(p<0.01\), *** \(p<0.001\).}
    \label{tab:kisub_main}
\end{table}

% ------------------------------------------------------------
% Treatment effects: AI substitution
% ------------------------------------------------------------

\FloatBarrier

\subsection{Human Oversight}
\label{subsec:hover_results}

\begin{figure}[htbp]
    \centering
    \includegraphics[
        width=\textwidth
    ]{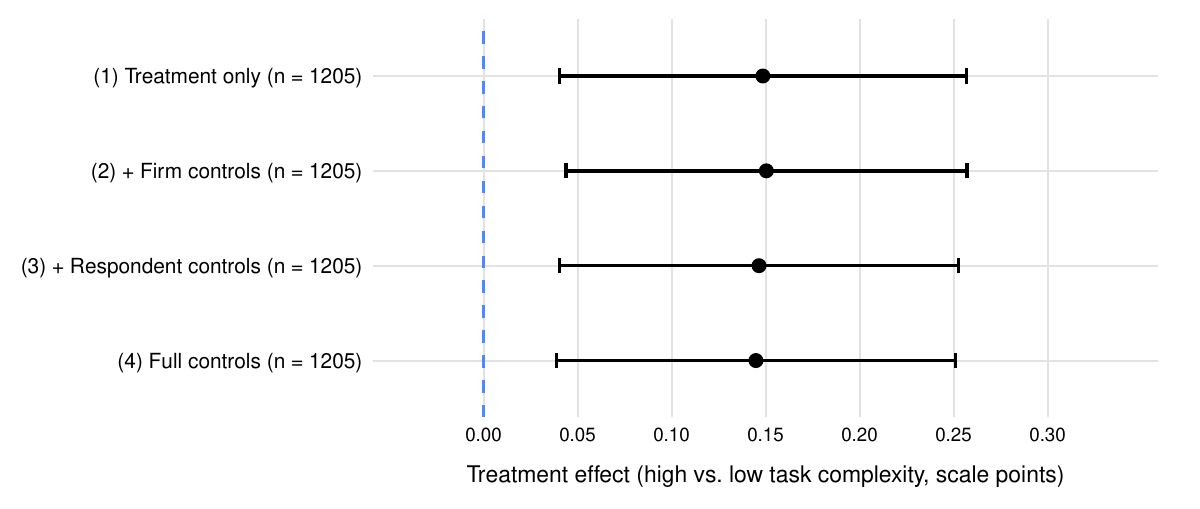}
    \caption{Effects of task complexity on the importance of human oversight}
    \label{fig:hover_treatment_effects}
    
    \vspace{0.2cm}
    \begin{minipage}{\textwidth}
        \footnotesize
        \textit{Notes:} Points show OLS estimates of the effect of high
        relative to low task complexity on the reported importance of human
        oversight. Horizontal lines indicate 95\% confidence intervals based
        on HC3-robust standard errors. All specifications are estimated using
        the common sample of the full model ($N=1{,}205$).
    \end{minipage}
\end{figure}

Accordingly, Table~\ref{tab:hover_main} reports the results of the nested OLS models for the assigned importance of human oversight. As before, column~(1) of Table~\ref{tab:hover_main} shows the coefficient $\beta_1$ without further controls. The estimation shows a positive effect of the high complexity condition on the importance that is assigned to human oversight. The effect is significant on the 1 percent level. Analogously to the previous models, adding firm and respondent characteristics as well as control variables regarding AI usage and information on previous hiring hardly changes the size of the coefficient of interest. Hence, when managers are presented with a scenario that involves more complex tasks, they assign a higher importance to critical judgement of AI generated outputs by a human. Our results therefore support Hypothesis~2. Figure~\ref{fig:hover_treatment_effects} shows the four estimators as well as their confidence bands graphically and displays that the estimation is robust across all four specifications.

The effect of the high complexity condition on human oversight score is considerably smaller than that on the substitution probability. In part because review is considered important even in the low-complexity condition (mean value for the assigned importance (scale 1 to 5) in the low complexity group: 4.32).

% Human oversight

\begin{table}[htbp]
    \centering
    
\begin{tabular*}{\textwidth}{@{\extracolsep{\fill}}lcccc}
\toprule
Variable & (1) & (2) & (3) & (4)\\
\midrule
High task complexity & 0.149** & 0.150** & 0.147** & 0.145**\\
 & (0.055) & (0.054) & (0.054) & (0.054)\\
AI applications used &  &  &  & 0.064**\\
 &  &  &  & (0.023)\\
Frequent personal AI use &  &  &  & 0.003\\
\addlinespace
 &  &  &  & (0.067)\\
Any previous entry-level hires: vocational &  &  &  & -0.155*\\
 &  &  &  & (0.067)\\
Any previous entry-level hires: university &  &  &  & 0.070\\
 &  &  &  & (0.065)\\
\addlinespace
Hiring difficulty: vocational &  &  &  & -0.029\\
 &  &  &  & (0.029)\\
Hiring difficulty: university &  &  &  & 0.012\\
 &  &  &  & (0.029)\\
\midrule
$R^2$ & 0.006 & 0.046 & 0.056 & 0.078\\
\addlinespace
Mean dep. var. (low complexity) & 4.323 & 4.323 & 4.323 & 4.323\\
Industry FE & No & Yes & Yes & Yes\\
Firm size FE & No & Yes & Yes & Yes\\
Respondent controls & No & No & Yes & Yes\\
Need indicators & No & No & No & Yes\\
\addlinespace
Observations & 1205 & 1205 & 1205 & 1205\\
\bottomrule
\end{tabular*}
    \caption{Effects of task complexity on the importance of human oversight. HC3-robust standard errors are reported in parentheses. Significance levels: + \(p<0.10\), * \(p<0.05\), ** \(p<0.01\), *** \(p<0.001\).}
    \label{tab:hover_main}
\end{table}

% ------------------------------------------------------------
% Treatment effects: Human oversight
% ------------------------------------------------------------

\FloatBarrier

\subsection{Expected Effects of AI on Hiring}
\label{subsec:hiring_expectations}

\begin{figure}[htbp]
    \centering
    \includegraphics[
        width=\textwidth
    ]{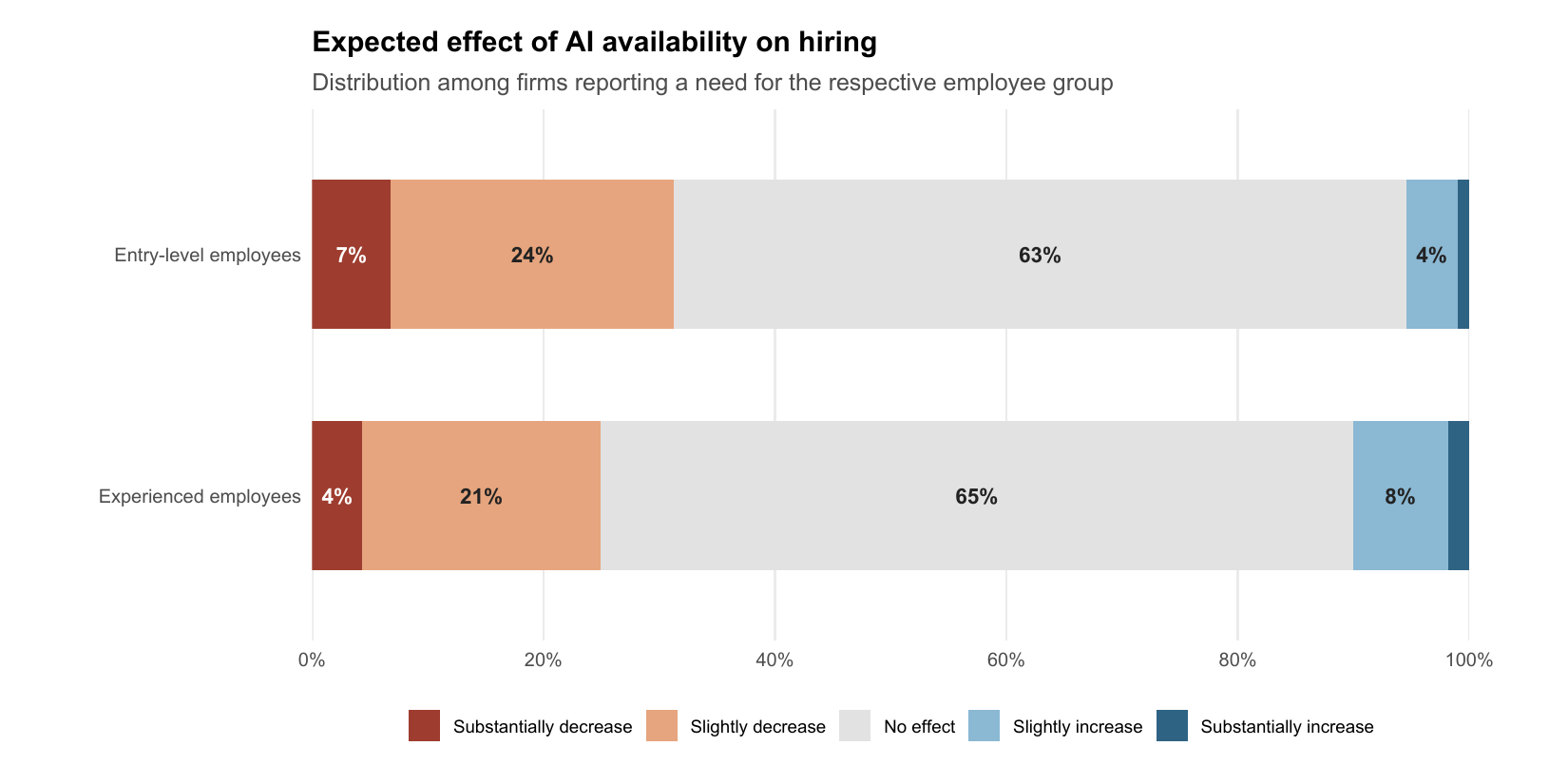}
    \caption{Expected effects of AI availability on hiring entry-level and experienced employees}
    \label{fig:expected_hiring_effects}
    
    \vspace{0.3em}
    \begin{minipage}{\textwidth}
        \footnotesize
        \textit{Notes:} The figure reports the distribution of managers' expected hiring responses to the availability of AI. The five-point Likert scale ranges from \textit{substantially decrease} to \textit{substantially increase}. Firms that (only) chose the outside options (i.e. report to have no need in general) are excluded.
    \end{minipage}
\end{figure}

The variables regarding hiring expectation provide a broader view of how managers believe AI availability will affect workforce demand over the next two years. Figure~\ref{fig:expected_hiring_effects} reports the response distribution among firms with a stated need for the respective employee group. The majority of firms do not expect a change in hiring behaviour: 63\% and 65\%, respectively, expect that the availability of AI applications has no effect on their hiring over the next two years. 

When looking at the response behaviour of those firms that expect AI availability to change their hiring behaviour, there are substantial differences between entry-level employees and experienced employees. For entry-level workers, 31.3\% expect hiring to decrease, whereas 5.5 per cent expect an increase. For experienced employees, the corresponding shares are 24.9\% that expect hiring to decrease, whereas 10.1\% expect an increase.  Accordingly, "no effect" is the modal response for both elicited groups, but negative hiring expectations are more common for entry-level hiring.

The mean expected effect for entry-level employees is 2.684 (95\% CI $[2.645,\,2.723]$), which is significantly below the neutral value of 3 ($t=-15.72$, $p<0.001$, $N=1{,}247$). This supports Hypothesis~3 that AI is generally seen as a substitute for entry-level work and that its availability might decrease hiring of these employees.\footnote{This conclusion should be interpreted with caution as our measures capture only the expected direction (increase / decrease of future hiring) rather than the number of additional or reduced hires. A larger share of firms expecting a general decrease in their hiring therefore does not necessarily imply a decline in aggregate hiring: in principle, this effect could be offset if the smaller group expecting an increase hires substantially more employees.}

For experienced employees we expected AI to be a complement, according to Hypothesis~4. Hence, we would expect the mean value for those employees to lie above the neutral value of three. However, the mean for experienced employees is also significantly below 3, at 2.826 (95\% CI $[2.788,\,2.865]$; $t=-8.76$, $p<0.001$, $N=1{,}291$). Hypothesis~4 is therefore not supported.

Finally, we test whether the hiring expectations for entry-level employees and experienced employees differ statistically significantly. Among the 1,230 firms that answered both items, the expected effect for experienced employees exceeds the expected effect for entry-level employees by 0.144 scale points (95\%  CI $[0.107,\,0.181]$, paired $t=7.56$, $p<0.001$). Appendix Table~\ref{tab:hiring_expectations_tests} reports the full test results. Hence, managers anticipate a relative shift towards experienced labour, but not as an absolute expansion of experienced hiring. 

\begin{figure}[htbp]
    \centering
    \includegraphics[
        width=\textwidth
    ]{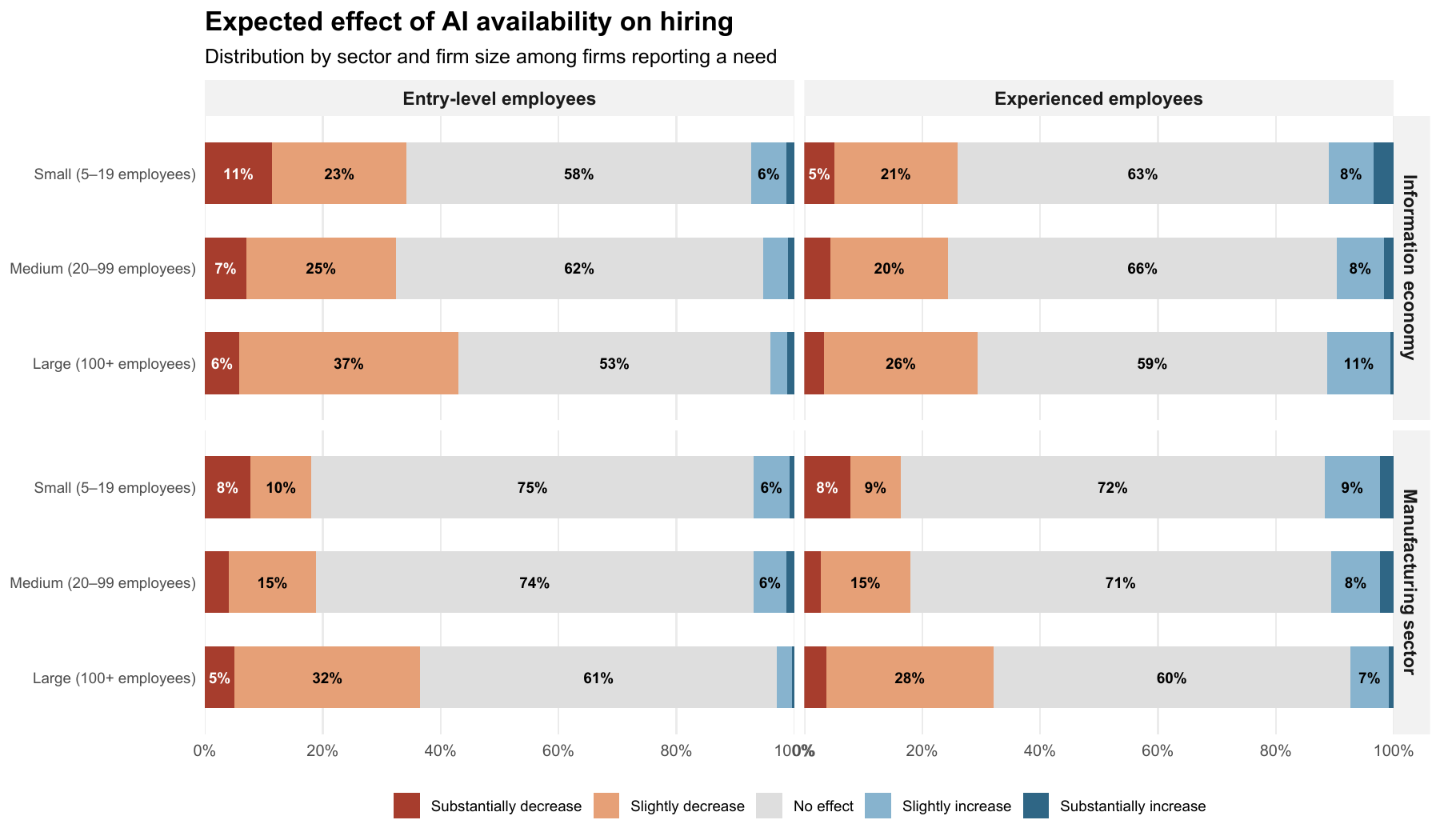}
    \caption{Expected effects of AI availability on hiring entry-level and experienced employees, by firm sector and size}
    
    \label{fig:hiring_sector_size}
    %    \vspace{0.2cm}
    \begin{minipage}{\textwidth}
        \footnotesize
        \textit{Notes:} The figure reports the distribution of expected hiring effects by sector, firm size, and employee group. Firms that (only) chose the outside options (i.e. report to have no need in general) are excluded.
    \end{minipage}

\end{figure}

When looking at firms' responses according to their sector and size, it becomes apparent that the expectations to decrease future hires are especially pronounced among large firms. As can be seen from Figure~\ref{fig:hiring_sector_size} 43\% of the large firms in the information economy expect to reduce their hiring behaviour regarding entry-level workers. That the share of firms expecting declining future hires is the largest among large firms with at least 100 employees is true for both industries and both types of employees. Similarly, the relative shift towards experienced workers can be seen across all firm characteristics.

\FloatBarrier

\FloatBarrier

\subsection{Hiring Expectations and Business Conditions}
\label{subsec:hiring_business_conditions}

Next, we examine whether managers' hiring expectations differ with their firms' recent and anticipated business conditions, i.e. their revenue dynamics. This analysis was not preregistered and should therefore be regarded as exploratory. The ZEW business survey always elicits whether the firms' revenue in the current quarter has risen, stayed equal or has declined compared to the previous quarter. In addition to the revenue development in the current (i.e. past as we send out the survey at the end of a quarter) quarter, they are asked about their revenue expectations for the next quarter. Again, the options are increase, equal, decrease.  We differentiate between three groups of firms that are mutually exclusive: (1) Firms fall into the category of having "\textit{sustained revenue growth}" if they state an increase in revenue for the previous quarter as well as report to expect increasing revenue for the next quarter. On the other side, firms that report a decline in revenue for the last quarter as well as further expected revenue declines in the next one are classified as experiencing \textit{sustained revenue decline}. All other firms with valid responses to both revenue questions are classified as having \textit{other revenue dynamics}.

\begin{figure}[htbp]
    \centering
    \includegraphics[width=\textwidth]{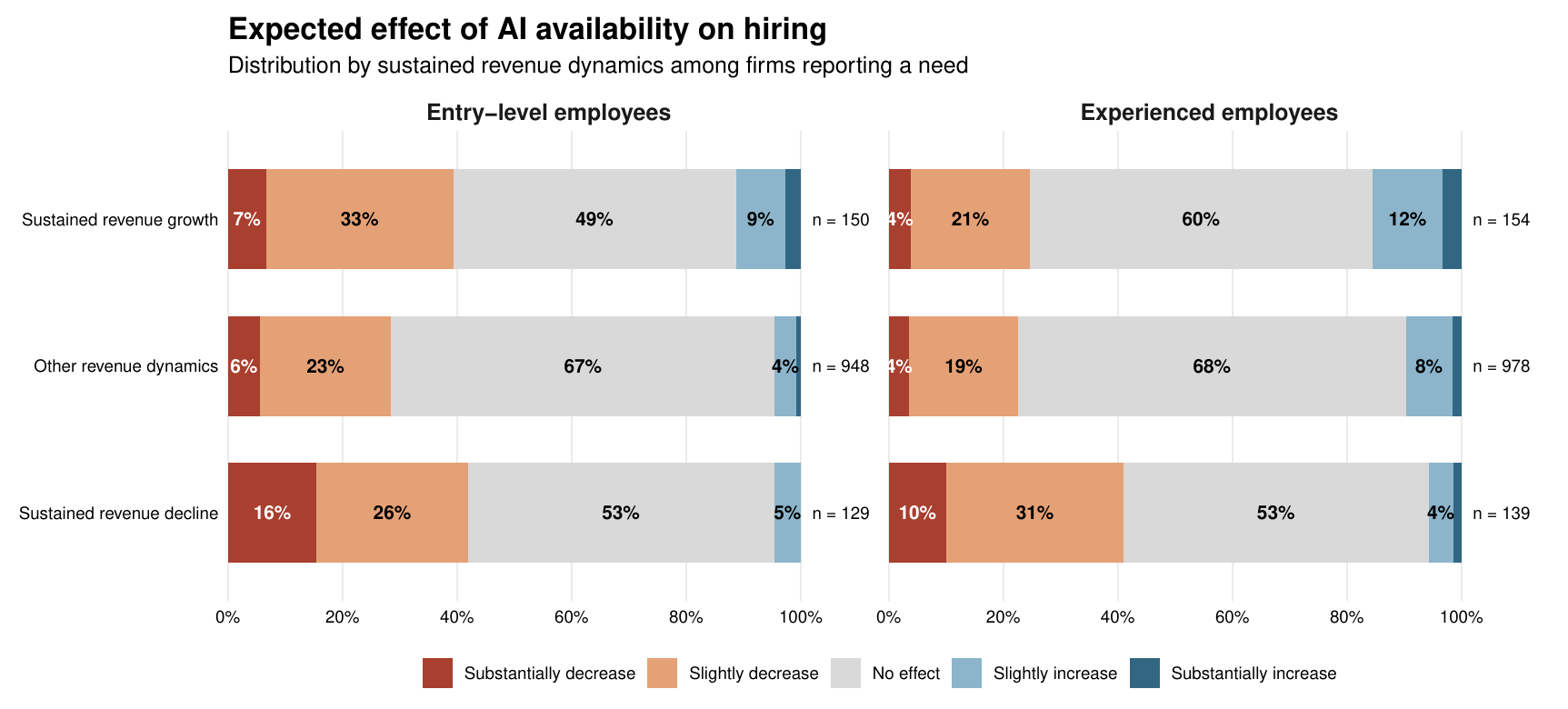}
    \caption{Expected effects of AI availability on hiring, by revenue dynamics}
    \label{fig:hiring_revenue_three_groups}
    \vspace{0.3em}
    \begin{minipage}{\textwidth}
        \footnotesize
        \textit{Notes:} The figure reports managers' expected effects of AI availability on hiring. Sustained revenue growth denotes firms reporting both an increase in revenue during the preceding quarter and an expected increase during the following quarter. Sustained revenue decline analogously requires decreases in both periods. Firms with all remaining combinations and valid responses are categorised as having other revenue dynamics. %Firms reporting no general need for the respective employee group are excluded. The number of valid observations is shown next to each bar. Percentages may not sum to 100 due to rounding. 
    \end{minipage}
\end{figure}

Figure~\ref{fig:hiring_revenue_three_groups} presents the complete response distributions for these three groups. As in the previous analyses, the samples are restricted to firms reporting a general need for the respective employee group. The entry-level sample comprises 150 firms experiencing sustained revenue growth, 948 firms with other revenue dynamics, and 129 firms experiencing sustained revenue decline. The corresponding numbers for experienced employees are 154, 978, and 139. Appendix Figure~\ref{fig:hiring_revenue_four_groups} provides a more detailed decomposition of the middle group.

\FloatBarrier

Firms that have sustained revenue decline also have substantially more negative hiring expectations. Regarding their expectations for hiring entry-level employees, 41.9\% of these firms expect the availability of AI tools to reduce their future hiring, whereas this share amounts to 28.5\% among firms with other revenue dynamics. There is also a visible difference when the lower end of the scale is regarded: 15.5\% of firms with sustained revenue decline expect substantial decreases in their future hiring of entry-level employees. Among the middle group the respective share is substantially smaller at 5.6\%. When asked about hiring experienced employees, 41\% of firms with declining revenue statements expect any decline (compared to 22.6\% within the group of firms with other revenue dynamics), 10.1\% even expect a substantial decrease in hiring experienced employees (compared to 3.6\% in the middle group). Hence, the share of firms expecting substantial decreases in hiring entry-level employees due to the availability of AI applications more than doubles among firms that had their revenue declining in the past quarter and expect a further decline in the next one. 

Comparisons based on the full five-point Likert scale further confirm this finding. The mean response on this scale is 2.473 for the hiring expectations for entry-level employees among firms experiencing sustained revenue decline, and 2.713 among firms with other revenue dynamics. The difference in means of \(-0.240\) scale points is statistically significant (HC3 SE \(=0.075\), 95\% CI [\(-0.387\), \(-0.094\)], \(p=0.001\)). Regarding the expectations of hiring experienced employees, the respective means amount to 2.561 and 2.852, yielding a difference of \(-0.291\) points (HC3 SE \(=0.071\), 95\% CI [\(-0.429\), \(-0.152\)], \(p<0.001\)). Both of these differences remain statistically significant after Holm-adjusting them due to the four group comparisons \citep{holm1979} (\(p_{\mathrm{Holm}}=0.004\) and \(p_{\mathrm{Holm}}<0.001\), respectively).

The differences are not driven by firm characteristics (i.e. firm size and sector). When industry and firm-size fixed effects are included, sustained revenue decline is associated with a reduction of 0.275 scale points in the entry-level outcome (HC3 SE \(=0.074\), \(p<0.001\)) and 0.304 points in the outcome for experienced employees (HC3 SE \(=0.072\), \(p<0.001\)). These estimates also remain significant after applying the Holm correction. Appendix Table~\ref{tab:revenue_mean_comparisons} reports all mean comparisons for both employee groups.

On the other side, the picture is more nuanced among firms with sustained revenue growth (i.e. firms that experienced an increase in revenue in the past quarter and expect their revenue to further increase in the next one). Those firms are descriptively more likely to expect an increase in their future hiring over the next two years due to AI availability. For entry-level employees, 11.3\% of firms with sustained growth expect an increase, compared with 4.5\% among firms with other revenue dynamics. For the experienced employees, the corresponding shares are 15.6\% and 9.7\%. In contrast, these firms are at the same time also more likely to expect a decrease, especially regarding entry-level hiring. Consequently, only 49\% of sustained-growth firms select "no effect" for entry-level employees, (compared with 67\% in the middle group). Hence, the response distribution is more strongly polarised in both directions for firms reporting sustained revenue growth, instead of being more optimistic overall. 

As a consequence, sustained revenue growth is not associated with a statistically significant difference in the mean response value. The difference between those firms and firms with other revenue dynamics is \(-0.033\) scale points for entry-level employees (\(p=0.643\)) and \(0.051\) points for experienced employees (\(p=0.447\)). When including industry and firm-size fixed effects, the corresponding estimates are \(-0.006\) and \(0.070\) and remain insignificant.

Direct binary comparisons provide suggestive evidence that firms with sustained revenue growth are more likely to expect hiring increases. For the entry-level employees, the difference of 6.8 percentage points of expecting any increase is statistically significant (\(p=0.012\)), but loses its significance at the 5\% level when we correct for multiple testing (\(p_{\mathrm{Holm}}=0.058\)). Regarding experienced employees, the difference amounts to  5.9 percentage points (\(p=0.058\); \(p_{\mathrm{Holm}}=0.173\)). There is no statistically significant difference in the probability of expecting a substantial increase. These patterns for the firms with sustained revenue growth, hence, have to be treated as descriptive rather than as clear evidence of an unambiguously positive shift in hiring expectations.

The binary comparisons reinforce the results for the firms experiencing a sustained revenue decline. After applying the Holm correction for multiple testing, the difference for expecting any decrease remains significant for both entry-level employees (\(p_{\mathrm{Holm}}=0.023\)) and experienced employees (\(p_{\mathrm{Holm}}<0.001\)). The difference in the share of firms expecting a substantial decrease is significant for entry-level employees (\(p_{\mathrm{Holm}}=0.019\)), but just falls outside the 5\% significance level for experienced employees (\(p_{\mathrm{Holm}}=0.058\)). Appendix Table~\ref{tab:revenue_share_comparisons} shows all eight comparisons, including the unadjusted and Holm-adjusted p-values.

Overall, the results indicate a robust association between sustained revenue decline and more negative expectations regarding future hiring due to the availability of AI applications. The results for sustained-growth firms instead suggest that managers in these firms are more likely to expect some adjustment in hiring, but disagree about its direction. These relationships should, however, not be interpreted as causal, as firms' revenue dynamics are not randomly assigned and current business conditions may capture broader differences in demand, investment, restructuring, or further managerial expectations.

\section{Additional Analyses and Robustness}
\label{sec:additional_analyses}

\subsection{Treatment Effect Heterogeneity}
\label{subsec:heterogeneity}

This section provides heterogeneity analyses. We examine whether the treatment effects vary with firm size, sector, firm-level AI use, frequent personal AI use, previous entry-level hiring, and recruitment difficulty. Appendix Figures~\ref{fig:kisub_heterogeneity} and~\ref{fig:hover_heterogeneity} report subgroup-specific effects estimated on the respective full-control sample and the 95\% confidence intervals. To assess heterogeneity formal interaction tests are used.

For AI substitution, the estimated treatment effect is negative in every analysed subgroup. The joint interaction tests do not reject equality by firm size ($p=0.430$), sector ($p=0.304$), firm AI use ($p=0.361$), personal AI use ($p=0.187$), previous entry-level hiring ($p=0.103$), or recruitment difficulty ($p=0.631$). The estimated treatment effects on human oversight are less precise across subgroups, as expected for the smaller average effect, but again show no statistically significant interaction.

Hence, our data provides no evidence that the treatment effect of higher task complexity is restricted to a specific group of firms or managers. The treatment effects remain visible accross the observed sample, but no systematic heterogeneity along the chosen explanatory variables was detected.

\subsection{Robustness Checks}
\label{subsec:robustness}

Our main specifications (OLS) treat the five-point outcome scales as numerical variables. We additionally estimate ordered probit models to account for their ordinal nature. As shown in Appendix Table~\ref{tab:probit}, the fully adjusted estimates are negative for AI substitution and positive for human oversight, with both treatment coefficients statistically significant at the 0.1 per cent level. The models use the same set of controls and the common samples as were used in the fully adjusted OLS specifications. Reported standard errors are heteroskedasticity-robust sandwich standard errors. The ordered probit coefficients refer to a latent index and cannot be directly compared to the OLS coefficients in size, but the direction and statistical significance of the treatment effects are consistent with the OLS findings.

Moreover, our results remain consistent when we exclude ambiguous responses, i.e. responses that combined scale values with the outside option. For instance, when a firm chose a value between 1 and 5 on how important they see human oversight of AI output, and simultaneously chose the outside option that they do generally not use AI, their provided scale value was used in our analyses. All results remain valid if these ambiguous responses are excluded. Further, our reported results used common sample specifications, i.e. all columns of Table~\ref{tab:kisub_main} and \ref{tab:hover_main} use the same sample, that is restricted by the most complex model in column (4) of both tables. Baseline results are not driven by this sample selection and remain stable in their size and significance if the full available sample is used for each model separately.

\section{Discussion}
\label{sec:discussion}

\subsection{Interpretation and Contributions}
\label{subsec:interpretation}

Our experiment isolates a feature of managers' decision-making that cannot be identified by observational data. The results reveal task complexity as an important factor of managers' willingness to use AI applications instead of hiring an entry-level employee. In both questionnaire conditions managers are presented with a scenario that assumes the long lasting need for additional capacity for tasks that are typically done by entry-level employees. The experimental variation alters the description of these tasks, especially their complexity. When the scenario includes more complex tasks that require professional judgement, contextual knowledge and independent problem-solving, managers are significantly less likely to use AI applications for these tasks instead of a new entry-level employee. Hence, the experiment shows that it is not the entry-level label itself that drives the perceived substitutability of work, but that this is dependent on the characteristics and the complexity of the tasks. 

Moreover, the task complexity treatment significantly affected the managers' assessment of the importance of human oversight of AI-generated output. Even though managers already attach great importance to human oversight when the tasks are well-structured and standardised, i.e. less complex, the desire for human oversight further increases when managers are asked about more complex tasks. This suggests that the task complexity does not only alter managers' willingness to use AI instead of human labour, but also the importance of human involvement and how necessary it is considered when tasks are offloaded to AI. 

Finally, the pre-treatment items on managers' hiring expectations over the next two years yield further insights on how AI might shape the labour markets. Managers expect the availability of AI applications to reduce hiring for both, entry-level employees as well as experienced employees. The expectations are significantly more negative for entry-level workers. Even though there is a relative difference between both career stages, our data does not support our hypothesis that AI might act as a substitute for entry-level work but as a complement to experienced work, as managers do not expect an absolute extension of future hiring for experienced employees. Yet, these responses are managers' expectations and should be treated as such. Real future outcomes and hiring developments might deviate from these stated expectations. 

The exploratory analyses of firms' revenue dynamics add further insights: Firms that report declining revenue for the previous quarter and simultaneously expect their revenue to decline further in the next quarter have more negative expectations regarding their future hiring activities due to AI. Firms that experienced recent revenue growth and expect further growth, on the other hand, are more likely to expect hiring increases, but are also more likely to expect decreases compared to firms that do not belong to one of those groups with extreme revenue developments. Hence, firms with bad business conditions report more negative hiring expectations due to AI, whereas firms with good business conditions give more polarised responses. 

For both firms as well as policy-makers a central question is whether potential automation of entry-level work might hinder professionals at the beginning of their careers to acquire the experience that is still wanted by firms. If standardised tasks provide a route into an occupation and hence to higher levels of experience, using AI instead of recruiting juniors could make that route more narrow. Training and work organisation should then find and provide other ways for new workforce entrants to develop the judgement skills and contextual knowledge that managers continue to value and demand.

%Extra für die Einreichung in Berlin:
%There are also implications for fiscal policy: Changes in access to employment and subsequent earnings may affect tax revenues and social-security contributions, while labour-market adjustment may create demand for publicly supported training and employment services. Our estimates do not quantify these fiscal effects or determine whether AI will increase or reduce net public revenue. They provide evidence on one hiring margin that is relevant to assessing such effects alongside productivity gains and the creation of new tasks.

\subsection{Limitations}
\label{subsec:limitations}

This study has several limitations. First, our measured outcomes are stated choices and expectations by managers and apply to a provided hypothetical scenario. The random assignment identifies how the task description changes the managers' likelihood of using AI instead of hiring a new entry-level employee or their desire for human oversight of AI-generated outputs, but it does not show how these firms would respond in reality. Hiring decisions depend on many other factors that we did neither provide nor observe, including firm strategy, wages and implementation costs, availability and quality of applicants, regulation and other variables that we cannot control in our setting. 

Second, we cannot rule out that different managers think about different tools when asked about "AI applications". We deliberately chose to keep the wording broad instead of naming a certain tool or technology, but this might lead to a diverse understanding among our respondents. Yet, our random assignment protects the treatment comparison as long as those variations of interpretation are balanced across both groups. 

Third, external validity is bounded by our sample. The data cover German firms in the information economy and the manufacturing sector. Results might be different for other sectors or countries. We observe that many firms have already had substantial exposure to AI applications before the survey. Other industries or economies where AI is less prominent might yield different results. Moreover, AI capabilities and firms' approaches to implementation are dynamic and rapidly changing. The managers' responses that are the foundation of the results stem from a survey that was conducted between June and August 2026 and should be treated as such.

Finally, the results regarding the expectations for future hiring behaviour are forecasts rather than realised changes in employment. It cannot be guaranteed that these forecasts will realise as expected. Moreover, we measure extensive margins rather than intensive margins. Although the data gives insights on how many firms expect their hiring activity to increase or decrease due to AI availability, it does not tell about the net effects in terms of how many candidates might be hired more or less. A large number of firms that decrease their hirings - but only slightly - might be offset by a small number of firms that increase their hiring activity by a lot.

\section{Conclusion}
\label{sec:conclusion}

This paper uses a preregistered survey experiment with managers at 1,435 German firms to investigate how task complexity affects their willingness to use AI instead of hiring an additional entry-level employee. Our setting allows us to experimentally vary task complexity in a hypothetical hiring scenario and elicit the likelihood of managers using AI applications instead of hiring additional entry-level employees as well as the importance they assign to human oversight of AI output. Our results support the pre-registered hypotheses that a higher level of task complexity leads to a lower likelihood of substituting entry-level labour with AI while it also increases the desire for critical human oversight. Both treatment coefficients have the expected sign and are statistically significant at the 1\% level in the common-sample OLS specifications. They remain statistically significant at least at the 5\% level across several robustness checks and are also supported by the ordered probit estimates.

Our results support the task-based view of technological change discussed earlier in the paper. Managers differentiate between different kinds of entry-level work, even when the given scenario including the entry-level label and the need for additional capacity are otherwise held constant. Standardised and structured tasks are more likely to be considered as suitable for substitution by AI, while complex tasks tend to maintain a stronger role for human judgement and review.

In addition to the two pre-registered hypotheses regarding the effect of the high-complexity setting on these post-treatment variables, we hypothesised that AI might be seen as a substitute to entry-level labour, but as a complement to experienced labour. 
% KI-Vorschlag fpr den Teil "While our data...towards experienced employees": Our data support the hypothesis that managers expect AI availability to reduce entry-level hiring. They do not support our prediction of increased hiring of experienced employees: the mean response also indicates an expected reduction for this group. Hiring expectations are, however, significantly less negative for experienced than for entry-level employees.
While our data confirm the hypothesis that AI acts as a substitute for junior work, we have to reject the hypothesis that AI and experienced employees act as complements, i.e. that the availability of AI tools may lead to an increase in firms' hiring expectations regarding more experienced employees. Instead, AI availability also decreases hiring expectations for experienced workers over the next two years. However, the expected reduction for them is significantly lower compared to the entry-level workers. Hence, we still see a relative shift in hirings towards experienced employees. We do not observe real labour market outcomes, but only managerial expectations for firms' future hiring behaviour. Exploratory analyses provide no statistically significant evidence that the treatment effect varies with firms' or managers' prior use of AI, reported recruitment difficulties, or previous entry-level hiring. Higher firm-level and personal AI use are nevertheless positively associated with managers' stated willingness to use AI applications instead of an additional entry-level hire.

Expectations to reduce hiring are especially pronounced among large firms and those who face economic challenges, i.e. suffer declining revenue and expect their revenue to decline further. Firms reporting and expecting revenue growth give more mixed responses, with no statistically significant difference in mean hiring expectations compared with the remaining revenue group, while descriptively exhibiting higher expectations of increasing hiring behaviour compared to firms with less prosperous revenue settings.

The overall sentiment of managers that the availability of AI tools will lead to a decrease in hiring -- not only but especially -- entry-level employees might have far-reaching consequences regarding both the labour market and the firms' own talent pipeline. When the willingness to hire entry-level workers declines, applicants without major expertise will have it much harder to enter the labour market. Additional trainings and labour market entry mechanisms might become necessary. On the other side, when firms require expertise in order to critically review AI-generated outputs, but refrain from hiring junior level workers, there might be a lack of experienced workers in the long run. Many firms rely on internal talent pipelines, growing their own experienced workforce. If this mechanism is disrupted, potentially for cost savings in the short run, firms may face difficulties developing the experienced workforce they need in the longer run. Managers as well as policy-makers should be aware of this conflict and the potential consequences firms might face in the future. A challenge for both is to realise and use the benefits that AI might provide for firms, but simultaneously offer routes through which new workforce entrants can acquire the experience and judgement skills that complex work continues to require.

\clearpage
\printbibliography
\clearpage

\appendix
\section*{Appendix}

\section{Survey Instrument and Experimental Materials}
\label{app:survey_instrument}

This appendix provides an English translation of the survey questions used in the present study. The original questionnaire was administered in German. Questions unrelated to the present study, including the regular business-cycle and working-from-home modules of the ZEW survey, are not reproduced here.

\subsection{Firm-Level and Personal AI Experience}
\label{app:ai_experience}

\begin{enumerate}[label=\textbf{Q\arabic*.}, leftmargin=*]

    \item \textbf{Which of the following artificial intelligence (AI) applications are used in your company?}

    For each application, respondents could select \textit{Yes} or \textit{No}.

    \begin{itemize}[leftmargin=*]
        \item Analysis or processing of text, speech, and image material
        \item Generative AI, for example for producing text or computer code
        \item Machine learning for data analysis and forecasting
        \item AI-supported workflow automation, for example invoice processing or customer-service chatbots
        \item Autonomous physical systems, for example warehouse and production robots or driverless transport systems
        \item Other AI applications
    \end{itemize}

    \item \textbf{Approximately how often do you personally use AI applications?}

    \begin{itemize}[leftmargin=*]
        \item Daily
        \item Several times per week
        \item Several times per month
        \item Approximately once per month
        \item Less than once per month
        \item Never
    \end{itemize}

\end{enumerate}

\subsection{Previous Entry-Level Hiring and Hiring Difficulties}

\begin{enumerate}[label=\textbf{Q\arabic*.}, leftmargin=*, resume]

    \item \textbf{Approximately how many entry-level employees has your company hired over the past two years?}

    Entry-level employees are defined as individuals taking up employment in their field for the first time, for example after completing vocational training or a university degree. This also includes retaining the company's own apprentices as employees. Respondents were asked to enter zero if no entry-level employees had been hired.

    \begin{itemize}[leftmargin=*]
        \item Number of entry-level employees with a vocational qualification: \underline{\hspace{2cm}}
        \item Number of entry-level employees with a university degree: \underline{\hspace{2cm}}
    \end{itemize}

    \item \textbf{Does your company generally experience difficulties finding and hiring entry-level employees?}

    Respondents answered separately for the following qualification groups:

    \begin{itemize}[leftmargin=*]
        \item Entry-level employees with a vocational qualification
        \item Entry-level employees with a university degree
    \end{itemize}

    \textbf{Response scale:}
    \(1=\) no difficulties;
    \(2\);
    \(3\);
    \(4\);
    \(5=\) major difficulties.

    \textbf{Outside option:} Generally no need.

\end{enumerate}

\subsection{Expected Effects of AI on Hiring}

\begin{enumerate}[label=\textbf{Q\arabic*.}, leftmargin=*, resume]

    \item \textbf{How do you think the availability of AI applications will affect the number of employees hired by your company over the next two years?}

    Respondents answered separately for:

    \begin{itemize}[leftmargin=*]
        \item Entry-level employees
        \item Experienced employees, that is, employees with sound knowledge in their field
    \end{itemize}
\newpage
    \textbf{Response options:}

    \begin{enumerate}[label=\arabic*., leftmargin=2em]
        \item Substantially decrease
        \item Slightly decrease
        \item No effect
        \item Slightly increase
        \item Substantially increase
    \end{enumerate}

    \textbf{Outside option:} Generally no need.

\end{enumerate}

\subsection{Experimental Manipulation}

Respondents were randomly assigned to one of the following two scenarios.

\subsubsection*{High Task Complexity}

\begin{quote}
Please imagine the following situation: Your company requires additional capacity in the long term for tasks that are typically performed by entry-level employees but require professional judgement, contextual knowledge, and independent problem-solving. Examples include interpreting information, preparing individualised recommendations, or handling more complex customer enquiries.
\end{quote}

\subsubsection*{Low Task Complexity}

\begin{quote}
Please imagine the following situation: Your company requires additional capacity in the long term for standardised, clearly structured tasks that are typically performed by entry-level employees. Examples include data maintenance, simple research, standardised analyses, or handling recurring customer enquiries.
\end{quote}

\subsection{Primary Experimental Outcomes}

Each respondent received the following questions immediately after the randomly assigned scenario.
\newpage
\begin{enumerate}[label=\textbf{Q\arabic*.}, leftmargin=*, resume]

    \item \textbf{How likely would your company be to use AI applications for such tasks instead of hiring an additional entry-level employee?}

    \textbf{Response scale:}
    \(1=\) very unlikely;
    \(2\);
    \(3\);
    \(4\);
    \(5=\) very likely.

    \textbf{Outside option:} We already do this.

    \item \textbf{How important would it be to you that AI-generated results for such tasks be critically reviewed by a human?}

    \textbf{Response scale:}
    \(1=\) not at all important;
    \(2\);
    \(3\);
    \(4\);
    \(5=\) very important.

    \textbf{Outside option:} We generally do not use AI.

\end{enumerate}

\subsection{Perceived Importance of Entry-Level Employees}

\begin{enumerate}[label=\textbf{Q\arabic*.}, leftmargin=*, resume]

    \item \textbf{How important are entry-level employees to your company for developing future skilled workers and managers?}

    \textbf{Response scale:}
    \(1=\) not at all important;
    \(2\);
    \(3\);
    \(4\);
    \(5=\) very important.

\end{enumerate}

\subsection{Firm and Respondent Characteristics}

\begin{enumerate}[label=\textbf{Q\arabic*.}, leftmargin=*, resume]

    \item \textbf{How many employees did your company have in Germany on average in 2025, including management and apprentices?}

    Number of employees: \underline{\hspace{2cm}}

    Share holding a university degree: \underline{\hspace{1.5cm}}\%

    \item \textbf{What is your age?}

    \begin{itemize}[leftmargin=*]
        \item Under 30
        \item 30--39
        \item 40--49
        \item 50--60
        \item Over 60
    \end{itemize}

    \item \textbf{What is your gender identity?}

    \begin{itemize}[leftmargin=*]
        \item Female
        \item Male
        \item Other
        \item Prefer not to answer
    \end{itemize}

\end{enumerate}

\clearpage
\section{Additional Tables and Figures}

%\subsection{Randomisation and Balance}

\FloatBarrier
\subsection{Alternative Estimators and Robustness Checks}

\begin{table}[htbp]
    \centering
    
\begin{tabular*}{\textwidth}{@{\extracolsep{\fill}}lcc}
\toprule
Variable & (1) & (2)\\
\midrule
Outcome & AI substitution & Human oversight\\
High task complexity & -0.501*** & 0.252***\\
 & (0.063) & (0.070)\\
\midrule
Industry FE & Yes & Yes\\
Firm size FE & Yes & Yes\\
\addlinespace
Respondent controls & Yes & Yes\\
AI experience controls & Yes & Yes\\
Previous hiring controls & Yes & Yes\\
Need and difficulty controls & Yes & Yes\\
Observations & 1210 & 1205\\
\bottomrule
\end{tabular*}
    \caption{Ordered probit estimates of the effects of task complexity on AI substitution and human oversight}
    \label{tab:probit}

    \vspace{0.2cm}
    \begin{minipage}{\textwidth}
        \footnotesize
        \textit{Notes:} The table reports ordered probit coefficients for the two main outcomes of interest, measured on five-point Likert scales. Both models include the complete set of controls that was used in the (fully adjusted) main models. Heteroskedasticity-robust sandwich standard errors are reported in parentheses. Ordered probit coefficients refer to the latent index and are not directly comparable with the OLS coefficients that are directly measured in scale points. Significance levels: \(+\ p<0.10\), \(*\ p<0.05\), \(**\ p<0.01\), \(***\ p<0.001\).
    \end{minipage}
\end{table}

\FloatBarrier
\clearpage
\subsection{Response Patterns and Outside Options}

\begin{table}[htbp]
    \centering
    \begin{threeparttable}
                \footnotesize
        \begin{tabular*}{\textwidth}{@{\extracolsep{\fill}}lcccccc}
\toprule
Response pattern & High complexity & Low complexity & Difference (pp) & HC3 SE (pp) & $p$-value & Observations \\
\midrule
\multicolumn{7}{l}{\textit{Panel A: AI substitution}} \\
\addlinespace[0.3em]
Scale only & 657 (93.2\%) & 668 (91.5\%) & 1.68 & 1.40 & 0.230 & 1435 \\
Outside option only & 18 (2.6\%) & 24 (3.3\%) & -0.73 & 0.89 & 0.409 & 1435 \\
Scale and outside option & 15 (2.1\%) & 31 (4.2\%) & -2.12 & 0.92 & 0.022 & 1435 \\
No answer & 15 (2.1\%) & 7 (1.0\%) & 1.17 & 0.65 & 0.074 & 1435 \\
Any outside-option selection & 33 (4.8\%) & 55 (7.6\%) & -2.82 & 1.28 & 0.027 & 1413 \\
Any valid scale response & 672 (95.3\%) & 699 (95.8\%) & -0.43 & 1.09 & 0.691 & 1435 \\
\midrule
\multicolumn{7}{l}{\textit{Panel B: Human oversight}} \\
\addlinespace[0.3em]
Scale only & 669 (94.9\%) & 679 (93.0\%) & 1.88 & 1.26 & 0.135 & 1435 \\
Outside option only & 23 (3.3\%) & 34 (4.7\%) & -1.40 & 1.03 & 0.175 & 1435 \\
Scale and outside option & 4 (0.6\%) & 7 (1.0\%) & -0.39 & 0.46 & 0.394 & 1435 \\
No answer & 9 (1.3\%) & 10 (1.4\%) & -0.09 & 0.60 & 0.877 & 1435 \\
Any outside-option selection & 27 (3.9\%) & 41 (5.7\%) & -1.82 & 1.13 & 0.110 & 1416 \\
Any valid scale response & 673 (95.5\%) & 686 (94.0\%) & 1.49 & 1.18 & 0.208 & 1435 \\
\bottomrule
\end{tabular*}

        \caption{Response patterns and outcome availability by treatment assignment}
        \label{tab:response_patterns}

        \begin{tablenotes}[flushleft]
            \footnotesize
            \item \textit{Notes:} Entries in the treatment columns (High / Low complexity) report the number (and percentages) of respondents exhibiting the respective response pattern. Differences are calculated as high task complexity minus low task complexity and are reported in percentage points. HC3 heteroskedasticity-robust standard errors are based on linear probability models. "Any outside-option selection" includes respondents who selected only the outside option or both the outside option and a valid scale value. "Any valid scale response" includes respondents who provided a scale value, including those who additionally selected an outside option.
        \end{tablenotes}
    \end{threeparttable}
\end{table}

\FloatBarrier
\subsection{Hiring Expectations}

\begin{table}[htbp]
    \centering
    \begin{threeparttable}
        \footnotesize
        \begin{tabular*}{\textwidth}{@{\extracolsep{\fill}}lrrrrrrr}
\toprule
Outcome or comparison & $N$ & Estimate & SD & 95\% CI & Reference & $t$-statistic & $p$-value \\
\midrule
\multicolumn{8}{l}{\textit{Panel A: Expected hiring effects by employee group}} \\
\addlinespace[0.3em]
Entry-level employees & 1247 & 2.684 & 0.710 & [2.645, 2.723] & 3 & -15.722 & $<0.001$ \\
Experienced employees & 1291 & 2.826 & 0.712 & [2.788, 2.865] & 3 & -8.762 & $<0.001$ \\
\midrule
\multicolumn{8}{l}{\textit{Panel B: Paired comparison between employee groups}} \\
\addlinespace[0.3em]
Experienced minus entry-level & 1230 & 0.144 & 0.667 & [0.107, 0.181] & 0 & 7.561 & $<0.001$ \\
\bottomrule
\end{tabular*}

        \caption{Expected effects of AI availability on hiring entry-level and experienced employees}
        \label{tab:hiring_expectations_tests}

        \begin{tablenotes}[flushleft]
            \footnotesize
            \item \textit{Notes:} Hiring expectations are measured on a five-point scale ranging from 1 (\textit{substantially decrease}) to 5 (\textit{substantially increase}); 3 indicates no expected effect.  Firms that (only) chose the outside options (i.e. report to have no need in general) are excluded. Panel A reports (two-sided) \(t\)-tests against the neutral scale value of 3. Panel B reports the difference between hiring expectations for entry-level employees and those for experienced employees (for firms that provided valid responses for both). 
        \end{tablenotes}
    \end{threeparttable}
\end{table}

\FloatBarrier

\newpage

\subsection{Detailed Revenue Dynamics}
\label{app:hiring_revenue_dynamics}

\begin{figure}[htbp]
    \centering
    \includegraphics[
        width=\textwidth,
        height=0.75\textheight,
        keepaspectratio
    ]{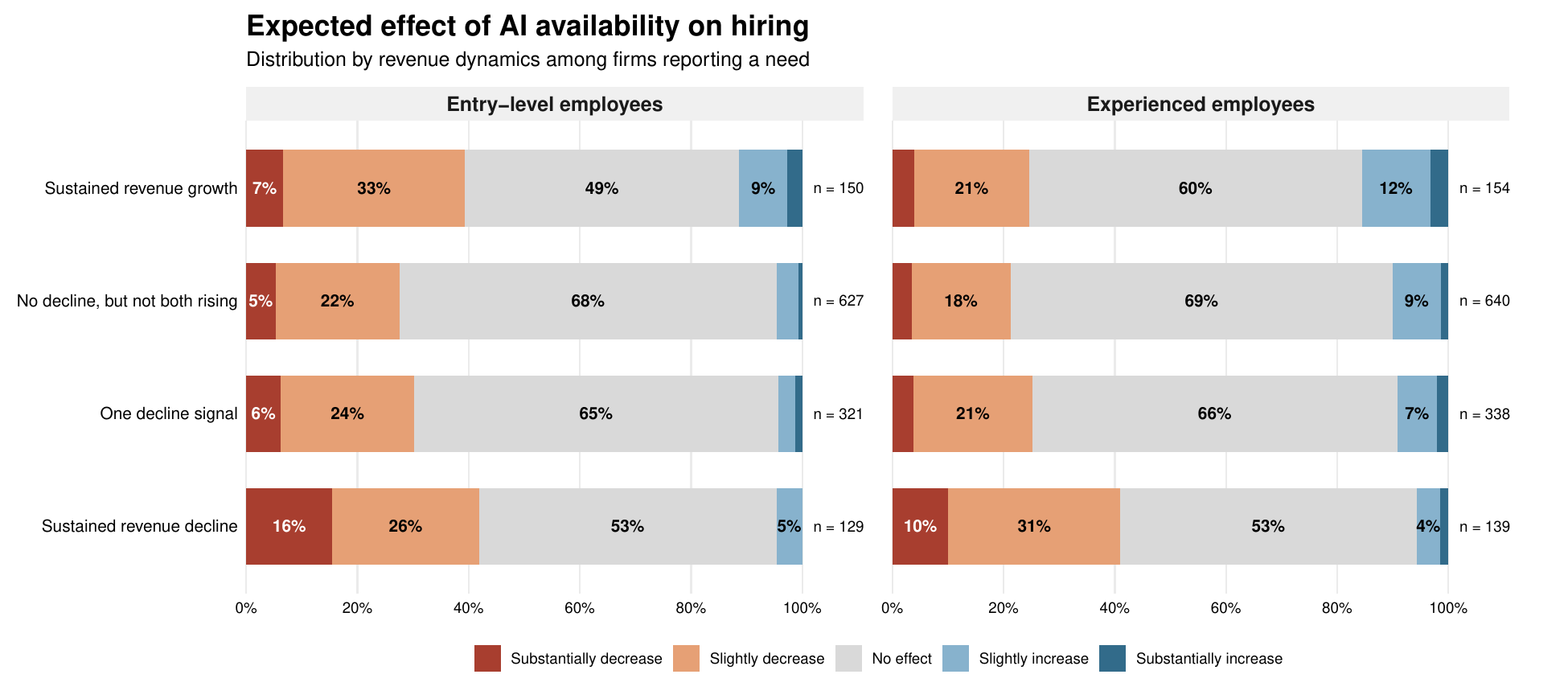}
    \caption{Expected effects of AI availability on hiring, by detailed revenue dynamics}
    \label{fig:hiring_revenue_four_groups}
    \vspace{0.3em}
    \begin{minipage}{\textwidth}
        \footnotesize
        \textit{Notes:} The figure reports managers' expected effects of AI availability on hiring across four (mutually exclusive) groups referring to the firms' recent and expected revenue development. Sustained revenue growth requires increases in both periods (i.e. the current/past quarter and the next quarter), whereas sustained revenue decline requires decreases in both periods. Other non-declining revenue dynamics comprise the remaining firms reporting no decrease in either period. One revenue decline signal denotes firms reporting a decrease in exactly one of both periods. 
    \end{minipage}
\end{figure}

Figure~\ref{fig:hiring_revenue_four_groups} provides a more detailed decomposition of the revenue classification used in Section~\ref{subsec:hiring_business_conditions}. The groups for sustained revenue growth and sustained revenue decline are identical to the main text. However, the former middle category from the main analysis is now divided into firms reporting no decline in either period (i.e. the current/past and next quarter) but do not report an increase in both periods, and firms reporting a decline in exactly one of the two periods. 

\FloatBarrier

% ------------------------------------------------------------
% Revenue dynamics: mean comparisons
% ------------------------------------------------------------

\begin{table}[htbp]
    \centering
    \begin{threeparttable}
        \footnotesize
        \setlength{\tabcolsep}{3pt}
        \begin{tabular*}{\textwidth}{@{\extracolsep{\fill}}llrrrrr@{}}
\toprule
Specification & Revenue group & N & Difference & HC3 SE & $p$ & $p_{\mathrm{Holm}}$ \\
\midrule
\multicolumn{7}{l}{\textit{Panel A: Entry-level employees}} \\
\addlinespace[0.3em]
Unadjusted & Sustained growth & 1227 & -0.033 & 0.071 & 0.643 & 0.893\\
Unadjusted & Sustained decline & 1227 & -0.240 & 0.075 & 0.001 & 0.004\\
Adjusted & Sustained growth & 1227 & -0.006 & 0.072 & 0.934 & 0.934\\
Adjusted & Sustained decline & 1227 & -0.275 & 0.074 & < 0.001 & < 0.001\\
\midrule
\multicolumn{7}{l}{\textit{Panel B: Experienced employees}} \\
\addlinespace[0.3em]
Unadjusted & Sustained growth & 1271 & 0.051 & 0.067 & 0.447 & 0.893\\
Unadjusted & Sustained decline & 1271 & -0.291 & 0.071 & < 0.001 & < 0.001\\
Adjusted & Sustained growth & 1271 & 0.070 & 0.067 & 0.297 & 0.594\\
Adjusted & Sustained decline & 1271 & -0.304 & 0.072 & < 0.001 & < 0.001\\
\bottomrule
\end{tabular*}
        \caption{Hiring expectations by revenue dynamics: differences in mean responses}
        \label{tab:revenue_mean_comparisons}
        \begin{tablenotes}[flushleft]
            \footnotesize
            \item \textit{Notes:} The table reports differences in mean hiring-expectation scores between each listed revenue group and firms with other revenue dynamics. Sustained growth denotes a firm that reports an increase in revenue for the current quarter compared to the previous one and simultaneously expecting a further increase in revenue for the next quarter; sustained decline analogously requires experienced and expected revenue decreases for both periods. The reference group comprises all remaining valid combinations. The measure for future hiring expectations range from 1 (substantially decrease) to 5 (substantially increase), with 3 indicating no (expected) effect. Adjusted models include industry and firm-size fixed effects. Standard errors are heteroskedasticity-robust (HC3). Holm-adjusted p-values jointly account for four comparisons within each specification: growth and decline relative to the reference group for each employee group. This analysis is exploratory.
        \end{tablenotes}
    \end{threeparttable}
\end{table}

% ------------------------------------------------------------
% Revenue dynamics: response-share comparisons
% ------------------------------------------------------------

\begin{table}[htbp]
    \centering
    \begin{threeparttable}
        \footnotesize
        \setlength{\tabcolsep}{2pt}
        \begin{tabular*}{\textwidth}{@{\extracolsep{\fill}}llrrrrrr@{}}
\toprule
Response & Revenue group & Other (\%) & Group (\%) & Diff. (pp) & SE (pp) & $p$ & $p_{\mathrm{Holm}}$ \\
\midrule
\multicolumn{8}{l}{\textit{Panel A: Entry-level employees}} \\
\addlinespace[0.3em]
Any decrease & Sustained decline & 28.5 & 41.9 & 13.4 & 4.6 & 0.004 & 0.023\\
Substantially decrease & Sustained decline & 5.6 & 15.5 & 9.9 & 3.3 & 0.003 & 0.019\\
Any increase & Sustained growth & 4.5 & 11.3 & 6.8 & 2.7 & 0.012 & 0.058\\
Substantially increase & Sustained growth & 0.8 & 2.7 & 1.8 & 1.4 & 0.180 & 0.359\\
\midrule
\multicolumn{8}{l}{\textit{Panel B: Experienced employees}} \\
\addlinespace[0.3em]
Any decrease & Sustained decline & 22.6 & 41.0 & 18.4 & 4.4 & < 0.001 & < 0.001\\
Substantially decrease & Sustained decline & 3.6 & 10.1 & 6.5 & 2.6 & 0.014 & 0.058\\
Any increase & Sustained growth & 9.7 & 15.6 & 5.9 & 3.1 & 0.058 & 0.173\\
Substantially increase & Sustained growth & 1.6 & 3.2 & 1.6 & 1.5 & 0.281 & 0.359\\
\bottomrule
\end{tabular*}
        \caption{Hiring expectations by revenue dynamics: differences in response shares}
        \label{tab:revenue_share_comparisons}
        \begin{tablenotes}[flushleft]
            \footnotesize
            \item \textit{Notes:} The values in the columns "Other" and "Group" report response shares among firms with other revenue dynamics and the listed revenue group, respectively. Differences are calculated as the listed group minus the reference group and are reported in percentage points (pp). Revenue groups are defined as in Table~\ref{tab:revenue_mean_comparisons}. "Any decrease" combines responses 1 and 2 (on the scale ranging from 1 to 5); "Substantially decrease" denotes response 1. "Any increase" combines responses 4 and 5; "Substantially increase" denotes response 5. Differences and HC3-robust standard errors are obtained from unadjusted linear probability models with revenue-group indicators. Holm-adjusted p-values account jointly for all eight comparisons reported in the table. The analysis is exploratory.
        \end{tablenotes}
    \end{threeparttable}
\end{table}

\FloatBarrier

\subsection{Treatment Effect Heterogeneity}
\begin{figure}[H]
    \centering
    \includegraphics[width=0.95\textwidth]{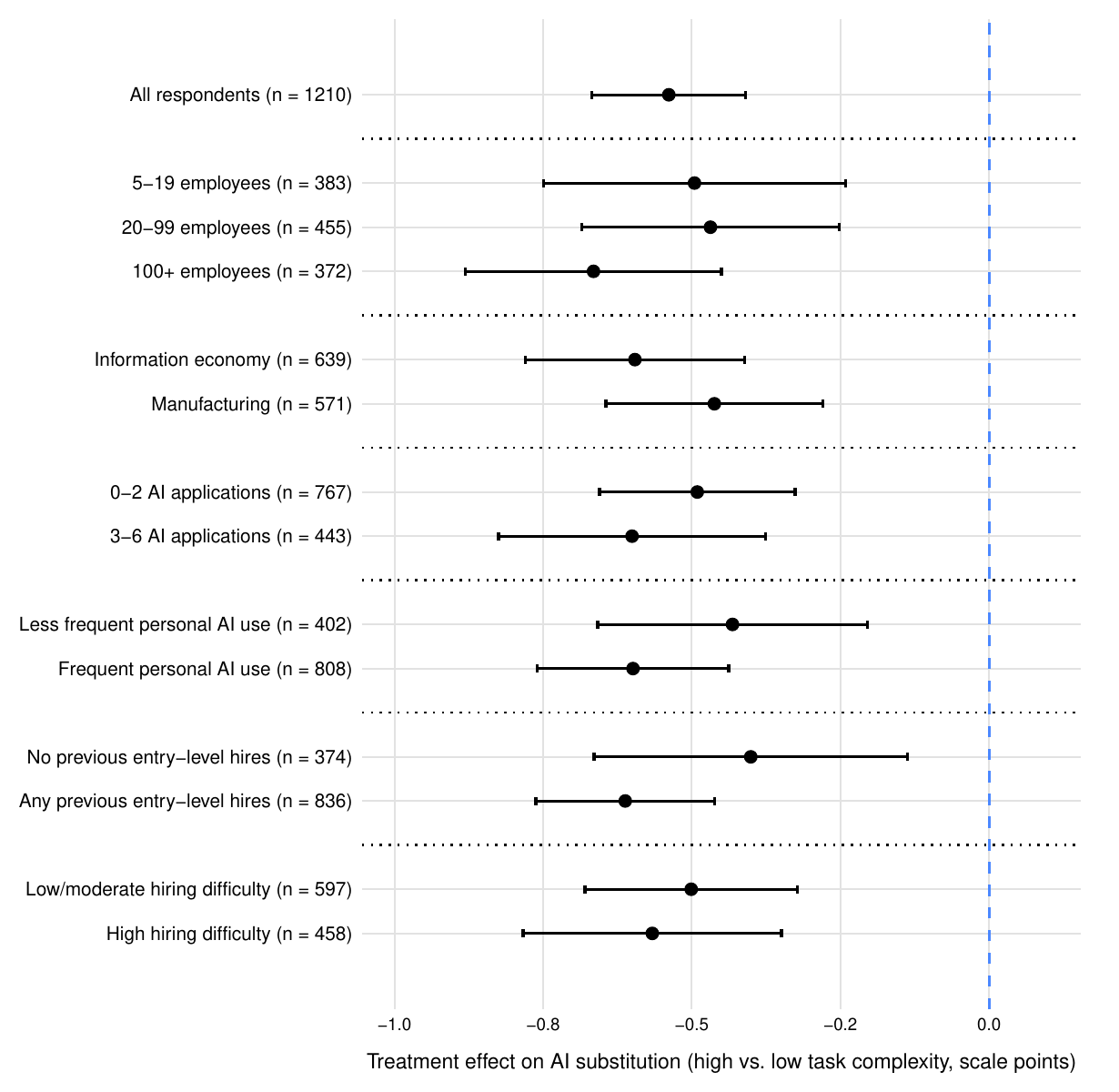}
    \caption{Treatment-effect heterogeneity in AI substitution}
    \label{fig:kisub_heterogeneity}

    \vspace{0.3em}
    \begin{minipage}{0.95\textwidth}
        \footnotesize
        \textit{Notes:} Points show subgroup-specific OLS estimates of the treatment effect (i.e. the effect of being in the high complexity setting) on willingness to use AI applications instead of hiring an additional entry-level employee. Horizontal lines represent 95\% confidence intervals, based on HC3 heteroskedasticity-robust standard errors. The analysis uses the complete-case sample of the fully adjusted main specification to increase comparability. Formal interaction tests provide no statistically significant evidence of treatment-effect heterogeneity across the displayed groups.
    \end{minipage}
\end{figure}

\begin{figure}[htbp]
    \centering
    \includegraphics[width=0.95\textwidth]{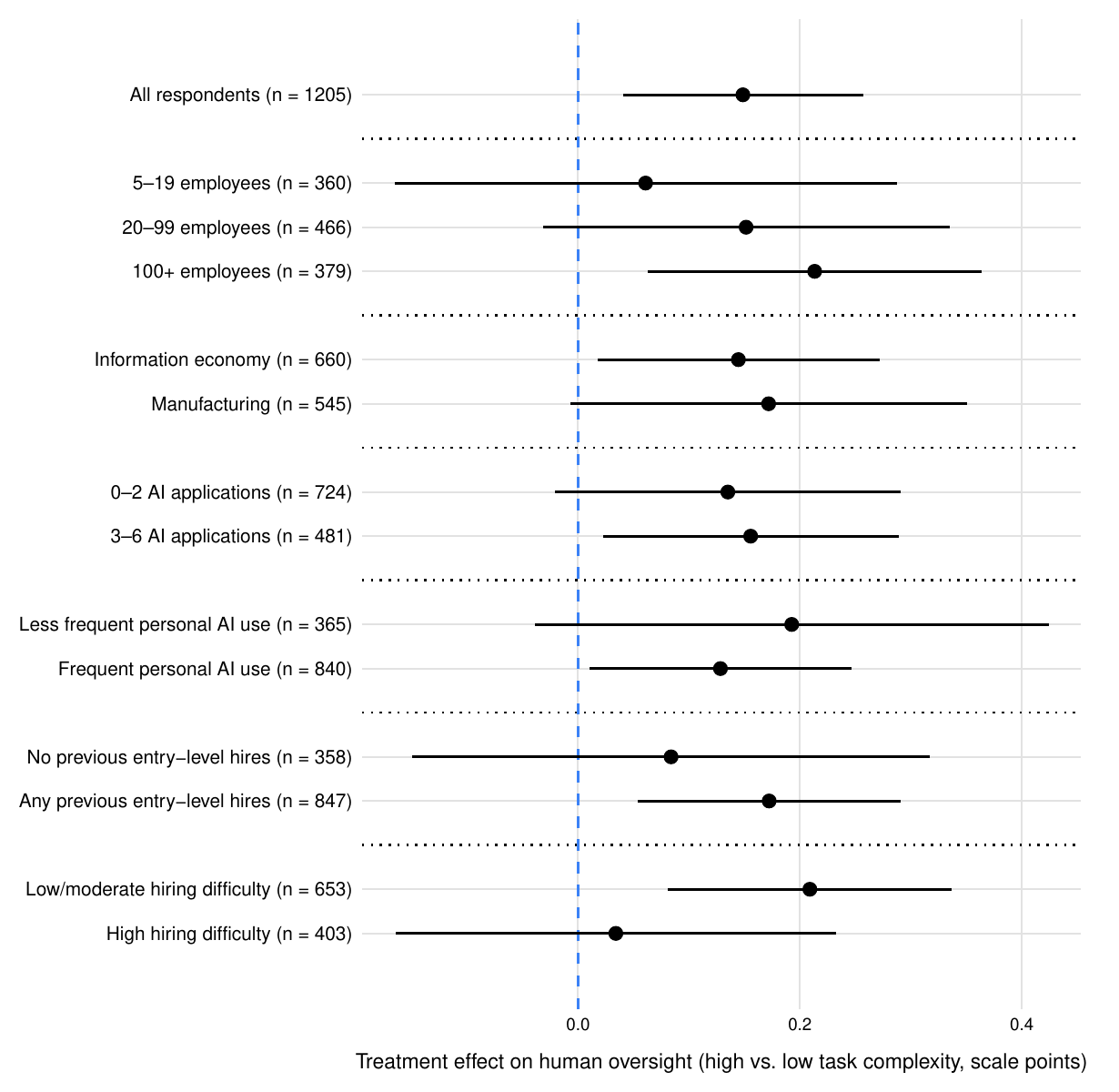}
    \caption{Treatment-effect heterogeneity in the importance of human oversight}
    \label{fig:hover_heterogeneity}

    \vspace{0.3em}
    \begin{minipage}{0.95\textwidth}
        \footnotesize
       \textit{Notes:} Points show subgroup-specific OLS estimates of the treatment effect (i.e. the effect of being in the high complexity setting) on the reported importance of human oversight for AI-generated output for the given set of tasks. Horizontal lines represent 95\% confidence intervals, based on HC3 heteroskedasticity-robust standard errors. The analysis uses the complete-case sample of the fully adjusted main specification to increase comparability. Formal interaction tests provide no statistically significant evidence of treatment-effect heterogeneity across the displayed groups.
    \end{minipage}
\end{figure}

\FloatBarrier
\subsection{Analytical Samples}

\begin{table}[htbp]
    \centering
    \begin{threeparttable}

        \small
        
\begin{tabular}{lrrrr}
\toprule
Sample & $N$ & Share of eligible sample (\%) & High complexity & Low complexity\\
\midrule
Eligible firm sample & 1435 & 100.0 & 705 & 730\\
Valid AI-substitution outcome & 1371 & 95.5 & 672 & 699\\
Full-control sample: AI substitution & 1210 & 84.3 & 584 & 626\\
Valid human-oversight outcome & 1359 & 94.7 & 673 & 686\\
Full-control sample: human oversight & 1205 & 84.0 & 588 & 617\\
\addlinespace
Valid values for both primary outcomes & 1309 & 91.2 & 647 & 662\\
Valid entry-level hiring expectation & 1247 & 86.9 & 618 & 629\\
Valid experienced hiring expectation & 1291 & 90.0 & 645 & 646\\
Valid hiring expectations for both groups & 1230 & 85.7 & 612 & 618\\
\bottomrule
\end{tabular}
        \caption{Analytical samples and treatment allocation}
        \label{tab:analytical_samples}

        \begin{tablenotes}[flushleft]
            \footnotesize
            \item \textit{Notes:} The eligible firm sample consists of firms with at least five employees that belong to the pre-defined industry sample frame (i.e. the German information economy or the German manufacturing industry). Percentages are calculated relative to the full eligible sample of 1,435 firms. Samples marked as "valid outcome" consist of respondents that provided a valid response on the respective five-point Likert scale. Samples denoted as "Full-control sample" contain the complete information set for all control variables that are included in the fully adjusted main models. The samples for the hiring expectation exclude firms that chose the outside option, i.e. reported to have generally no need for the respective employee group. "High complexity" and "Low complexity" refer to the two (randomly assigned) treatment conditions (i.e. scenario descriptions) of the experiment.
        \end{tablenotes}
    \end{threeparttable}
\end{table}

\end{document}